\documentclass[]{aa} 

\usepackage{graphicx}
\usepackage{txfonts}
\usepackage{color}
 \def\mean#1{\left< #1 \right>}
\usepackage[]{hyperref}
\usepackage{ gensymb }

\begin{document}

        \title{The soft excess of the NLS1 galaxy Mrk 359 studied with an \textit{XMM-Newton}-\textit{NuSTAR} monitoring campaign}
        
        \subtitle{}
        
  \author{R. Middei
        \inst{1,2,3}\fnmsep\thanks{riccardo.middei@ssdc.asi.it}
        \and P.-O. Petrucci \inst{4} \and S. Bianchi\inst{1} \and F. Ursini \inst{1} \and M. Cappi \inst{5} \and M. Clavel  \inst{4} \and \\ A. De Rosa \inst{6} \and A. Marinucci \inst{7} \and G. Matt \inst{1} \and A. Tortosa \inst{8}
}
\institute{Dipartimento di Matematica e Fisica, Universit\`a degli Studi Roma Tre, via della Vasca Navale 84, I-00146 Roma, Italy
        \and Space Science Data Center, SSDC, ASI, via del Politecnico snc, 00133 Roma, Italy
        \and INAF - Osservatorio Astronomico di Roma, via Frascati 33, I-00040 Monteporzio Catone, Italy
        \and
        Univ. Grenoble Alpes, CNRS, IPAG, F-38000 Grenoble, France.
\and
INAF-Osservatorio di astrofisica e scienza dello spazio di Bologna, Via Piero Gobetti 93/3, 40129 Bologna, Italy.
\and
INAF/Istituto di Astrofisica e Planetologia Spaziali, via Fosso del Cavaliere, 00133 Roma, Italy.
\and ASI – Unit\`a di Ricerca Scientifica, Via del Politecnico snc, I-00133 Roma, Italy
\and
N\'ucleo di Astronom\'ia de la Facultad dei Ingenier\'ia, Universitad Diego Portales, Av. Ej\'ercito Libertador 441, Santiago, Chile.
}

        
        \abstract
        {Joint \textit{XMM-Newton} and \textit{NuSTAR} multiple exposures allow us to disentangle the different emission components of active galactic nuclei (AGNs) and to study the evolution of their different spectral features. In this work, we present the timing and spectral properties of five simultaneous \textit{\textit{XMM-Newton} \emph{and} \textit{NuSTAR}} observations of the Narrow Line Seyfert 1 galaxy Mrk 359.}
        {We aim to provide the first broadband spectral modeling of Mrk 359 describing its emission spectrum from the UV up to the hard X-rays.}
        {We performed temporal and spectral data analysis, characterising the amplitude and spectral changes of the Mrk 359 time series and computing the 2-10 keV normalised excess variance. The spectral broadband modelling assumes the standard hot Comptonising corona and reflection component, while for the soft excess we tested two different models: a warm, optically thick Comptonising corona (the two-corona model) and a reflection model in which the soft-excess is the result of a blurred reflected continuum and line emission (the reflection model).}
        {High and low flux states were observed during the campaign. The former state has a softer spectral shape, while the latter shows a harder one. The photon index is in the 1.75--1.89 range, and only a lower limit to the hot-corona electron temperature can be found. A constant reflection component, likely associated with distant matter, is observed. Regarding the soft excess, we found that among the reflection models we tested, the one providing the better fit (reduced $\chi^2$=1.14) is the high-density one. However, a significantly better fit (reduced $\chi^2$=1.08) is found by modelling the soft excess with a warm Comptonisation model.}
        {The present analysis suggests the two-corona model as the best scenario for the optical-UV to X-ray emission spectrum of Mrk 359.}
        
        
        \maketitle
\section{Introduction}

The primary X-ray emission of active galactic nuclei (AGNs) is usually explained by invoking a two-phase scenario \citep[e.g.][]{haar91,haar93,haar94} in which thermal electrons (the so-called hot corona) intercept and Comptonise optical-UV photons arising from the accretion disc. Different arguments \citep[timing and microlensing, e.g.][]{Chartas09,Morgan12,DeMa13,Uttley14,Kara16} support the hot corona to ensure it remains compact and located in the inner regions of the accretion flow. The X-ray primary spectrum is well approximated by a cut-off power law with a photon index ($\Gamma_{\rm hc}$) in the range of 1.5-2.5 \citep[e.g.][]{Bian09}, and the high-energy rollover (E$_{\rm{c}}$) is interpreted as a further signature of thermal Comptonisation. Such phenomenological quantities depend on the opacity and temperature of the hot corona \citep[e.g.][]{Shapiro1976,Rybi79,Sunyaev1980,Belo99,Petrucci01,Middei2019}.\\
\indent Circumnuclear matter can reprocess part of the X-ray continuum, giving rise to an Fe K$\alpha$ fluorescent emission line \citep[e.g.][]{Bian09}, which is sometimes accompanied by weaker features due to ionised iron \citep[see e.g.][]{Molendi2003,Bianchi2005}. The neutral Fe K$\alpha$ emission line has an intrinsically narrow profile. However, Doppler shift and gravitational redshift (due to special and general relativity, respectively) can broaden the line profile, and this effect is stronger when the reflection materials are closer to the supermassive black hole \citep[SMBH, e.g.][]{Fabi00,Matt93}. Additionally, reflection off Compton thick matter gives rise to the so-called Compton hump at about 30 keV \citep[e.g.][]{George91,Matt91,Matt93,Garc14a,Daus16}.\\
\indent An emission in excess of the extrapolated high-energy power law \citep[e.g.][]{Bian09,Gliozzi2019} is often observed in the soft part (below $\sim$1-2 keV) of AGNs' X-ray spectra. The origin of this soft excess is still debated, and different competing models have been proposed to explain it. At present, the two most popular ones are blurred ionised reflection from the innermost regions of the accretion disc \citep[e.g.][]{Miniutti04,Crummy06,Walton2013,Liebmann2018}, and a warm (kT$\sim$0.5-1 keV), optically thick Comptonising layer above the accretion disc \citep[e.g.][]{Magdziarz1998,Jin2012,Done2012,Petrucci2013,Kubota2018,Petrucci18,Petrucci2020}.\\
\object{The Mrk~359 galaxy} \citep[$z$=0.0174,][]{Maurogordato97} is the first discovered narrow-line Seyfert galaxy (NLS1) (\cite{Davi78}, \citet{Osterbrock83}, where the width of the broad emission line is smaller than 2000 km/s. \cite{Elvis92} first reported on the X-ray properties of this source in the context of the Einstein Slew Survey. \cite{Boller96} used \textit{ROSAT} data to report on the steep soft X-ray emission of Mrk 359 ($\Gamma$=2.4$\pm$0.1), while \cite{Walter93} claimed the presence of a strong soft X-ray excess. The best knowledge of the X-ray behaviour of this source is derived from an archival \textit{XMM-Newton} observation (July 2000) analysed by \citet{OBrien01}. The authors found a power-law-like spectrum with $\Gamma\sim1.84$ in the 2-10 keV band, a prominent soft X-ray excess and no significant neutral/warm intrinsic absorption. The spectrum showed a noticeable neutral iron emission line (EW$\sim$200 eV) modelled via two components of approximately equal strengths: a broad iron line from an accretion disc and a narrow unresolved line core (EW$\sim$120 eV) consistent with fluorescence from neutral iron in distant reprocessing matter. This AGN was later analysed by \cite{Walton2013}, who studied a 2007 \textit{Suzaku} observation. They modelled the source spectrum with a power-law-like primary continuum and a reflection component computed using \textit{REFLIONX} \citep{Ross2005}. \cite{Walton2013} found the primary photon index to be $\Gamma$=1.90$\pm$0.03, and explained the soft excess with a blurred ionised reflection.\\
\indent  In this paper, we report on the \textit{XMM-Newton-NuSTAR} observational campaign targeting the unobscured AGN. The aim of the campaign was to study the broadband, time-resolved spectra of this source and shed light on the nature of its prominent soft X-ray excess. The standard cosmology \textit{$\Lambda$CDM} with H$_0$=70 km/s/Mpc, $\Omega_m$=0.27 and $\Omega_\lambda$=0.73 is adopted throughout this work.
\section{Data reduction}
\begin{figure}
        \centering
        \includegraphics[width=0.49\textwidth]{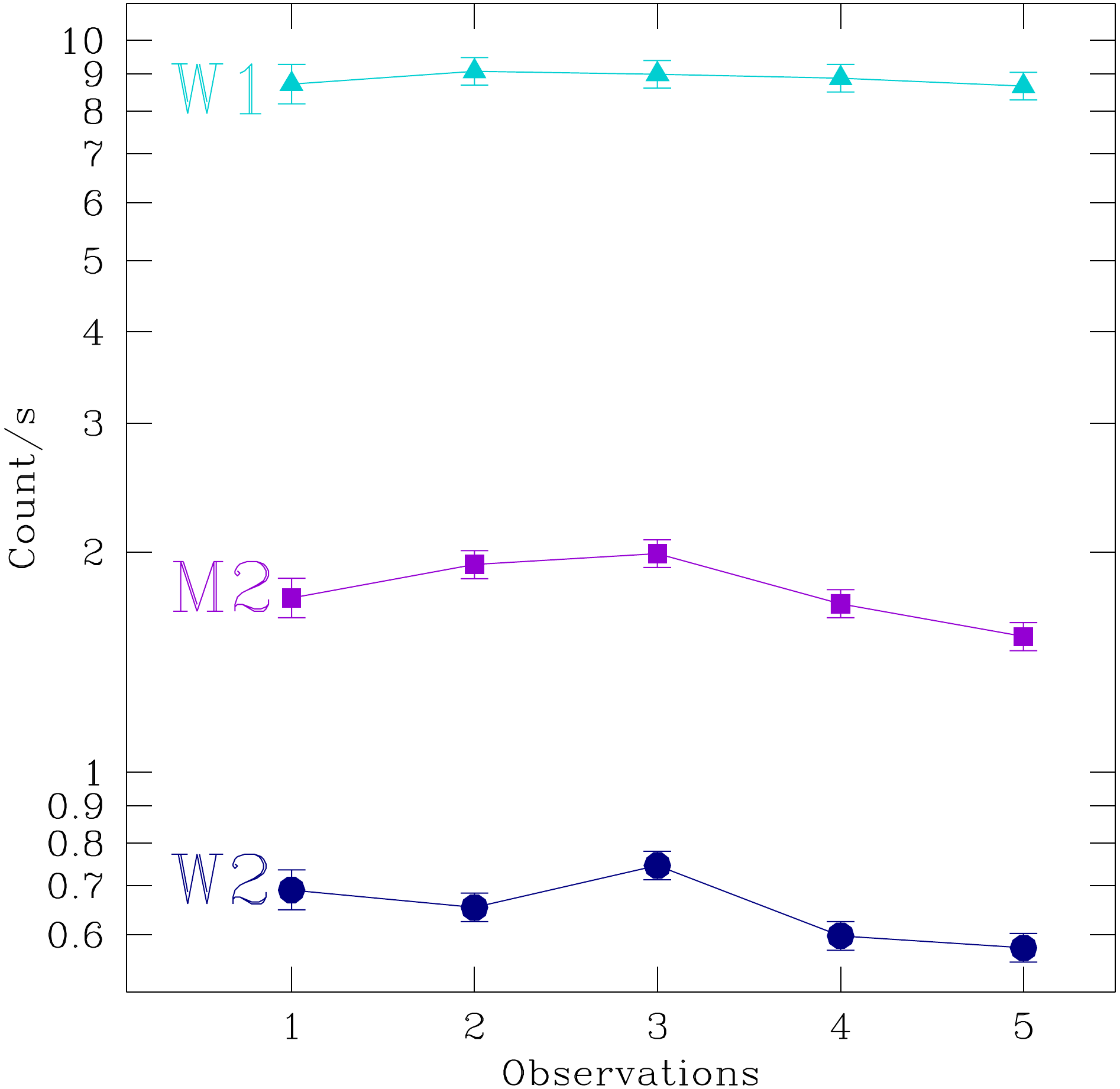}         
        \caption{\small{\textit{OM} light curves for the five observations of the 2019 campaign. The different filters are labelled in the plot. \label{om}}}
\end{figure}
\begin{figure}
        \centering      
        \includegraphics[width=0.48\textwidth]{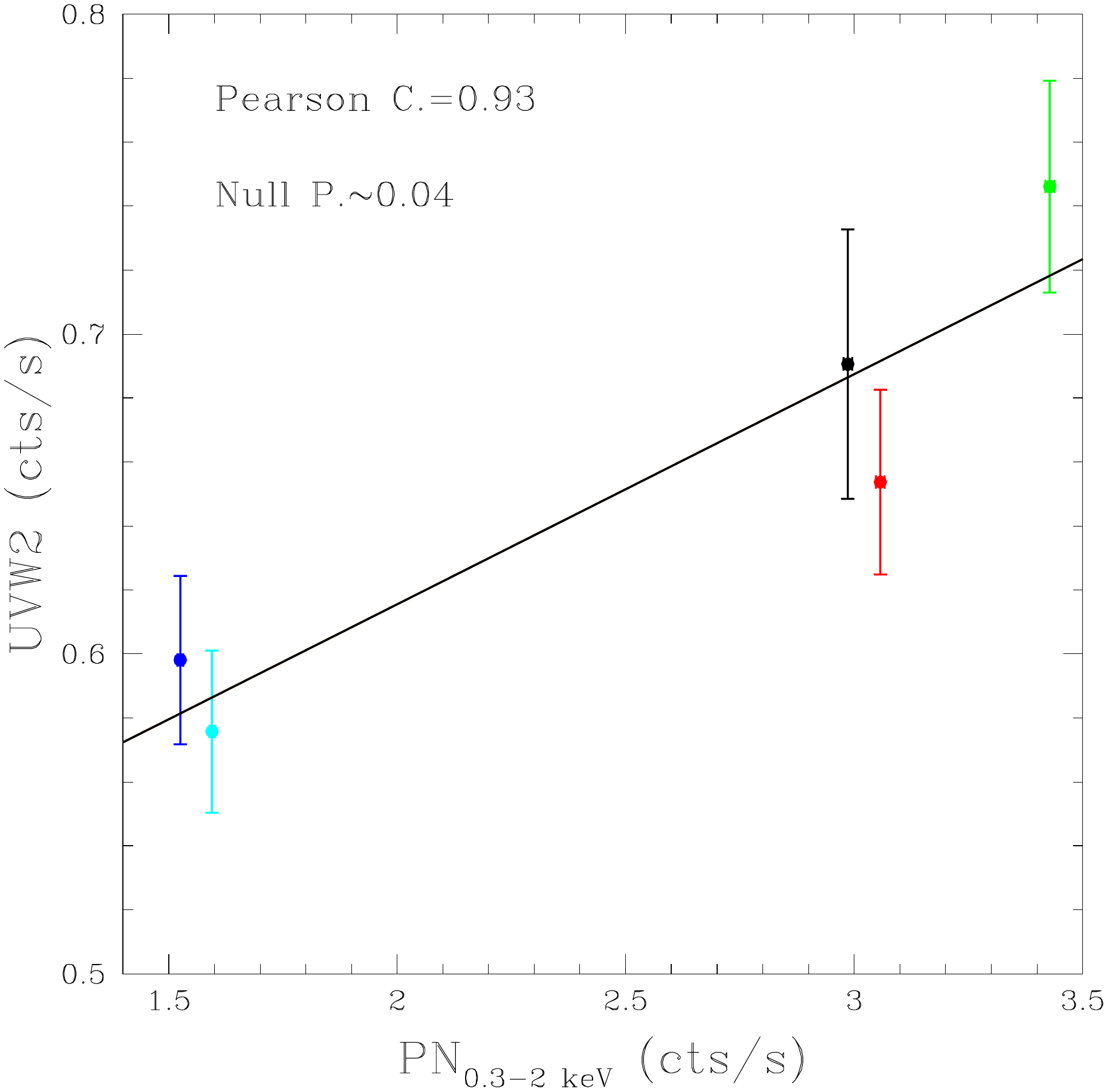} 
        \caption{\small{UVW2 rates are shown as a function of the \textit{EPIC-pn} count rate in the 0.3-2 keV band. The suggestive correlation between these two quantities is supported by the Pearson correlation coefficient and its low null probability reported in the graph.}}
\end{figure}
\begin{figure}
        \centering
        \includegraphics[width=0.49\textwidth]{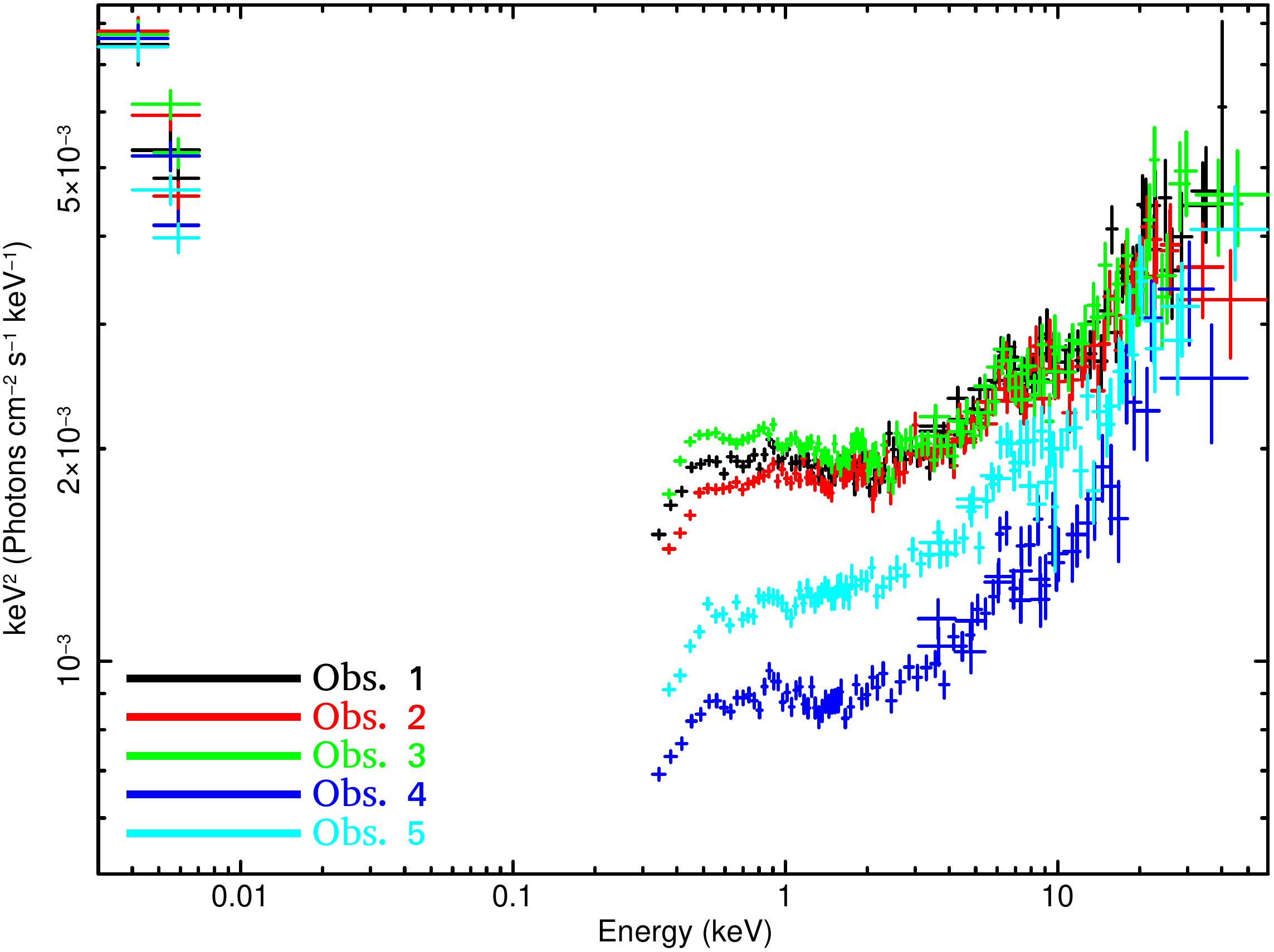}    
        \caption{\small{Unfolded spectra ($\Gamma=2$) as observed by \textit{XMM-Newton} (\textit{EPIC-pn} and \textit{OM}) and \textit{NuSTAR} are presented. The colours black, red, green, blue, and cyan account for visit 1, 2, 3, 4, and 5, respectively. This colour code is applied throughout the whole paper.\label{unfo}}}
\end{figure}

The present dataset consists of five \textit{XMM-Newton-NuSTAR} simultaneous exposures taken at the beginning of 2019 (see Table \ref{log}). Analyses were performed on \textit{XMM-Newton-pn/OM} and \textit{NuSTAR FPMA/B} data, while the \textit{MOS} detectors were not considered due to the lower statistics of their spectra.\\
\indent \textit{\emph{The} \textit{XMM-Newton}} data were processed using the \textit{XMM-Newton} Science Analysis System ($SAS$, Version 18.0.0). The choice of the source extraction radius and the screening for high background time intervals were performed by an iterative process aimed at maximising the signal-to-noise ratio \citep[S/N, details in][]{Pico04}. For each observation, we obtain the source radius that maximises the S/N ratio. These radii span between 15 and 40 arcsec, where smaller values are used when the background is higher (specifically for observations 1 and 5). At this stage it is worth noting that these different sizes for the extracting regions are taken into account during the computations of the spectra, hence the results of the spectral analysis are not affected. The background was extracted using a 40 arcsec region close to the source. Spectra were binned to have at least 30 counts for each spectral bin, and not to oversample the instrumental energy resolution by a factor larger than three. We notice that no significant pile-up affects the pn data, as also indicated by the $SAS$ task $epatplot$. Data obtained with the Optical Monitor \citep[][\textit{OM}]{Maso01} are also used. This telescope on-board \textit{XMM-Newton} observed Mrk 359 in the UVM2 (2310 \AA), UVW1 (2910 \AA), and UVW2 (2120 \AA) filters, and such measurements are available for all the visits. \textit{\emph{The} OM} data were extracted using the on-the-fly data processing and analysis tool \textit{RISA}, the remote interface SAS analysis and spectral points were converted into a convenient format to be analysed with \textit{Xspec} \cite[][]{Arna96} using the standard task \textit{om2pha}. \textit{\emph{The} OM} light curves for the different filters are reported in Fig~\ref{om}.\\
In Fig.~\ref{Xuv}, we show the \textit{XMM-Newton}/pn average count rates for each observation versus the UVW2 count rate. The correlation between the 0.3-2 keV band and the UVW2 is supported by a Pearson's correlation coefficient of 0.93 accompanied by a null hypothesis probability of 0.04. \\
\indent \textit{\emph{The} NuSTAR} observations were reduced using the standard pipeline ($nupipeline$) in the \textit{NuSTAR} Data Analysis Software and by adopting the calibration database x20180710. High-level products were obtained using the $nuproducts$ routine for both the hard X-ray detectors \textit{FPMA/B}. A 40-arcsec-radius circular region was used to extract the source spectra, while the background was computed from a blank area with the same radius near the source. \textit{NuSTAR} spectra were then binned to have an S/N greater than three in each spectral channel, and not to oversample the energy resolution by a factor larger than 2.5. The spectra of the whole dataset, unfolded using a $\Gamma=2$ and a common unitary normalisation, are shown in Fig.\ref{unfo}.

\begin{table}
        \centering
        \caption{\small{Satellite, observation ID, start date, and net exposure time are reported.}\label{log}}
        \begin{tabular}{c c c c}
                \hline
                Observatory & Obs. ID & Start date & Net exp. \\
                & & yyyy-mm-dd & ks \\
                \hline
                \hline
                \textit{XMM-Newton} & 0830550801 &2019-01-25 & 35\\
                \textit{NuSTAR} & 60402021002 &  & 52 \\
                \hline
                \textit{XMM-Newton} & 0830550901 & 2019-01-26  & 37\\
                \textit{NuSTAR} & 60402021004 &  & 49 \\
                \hline
                \textit{XMM-Newton} & 0830551001 &2019-01-28 & 35 \\
                \textit{NuSTAR} & 60402021006 && 51 \\
                \hline
                \textit{XMM-Newton} &0830551101  &2019-01-31  & 40 \\
                \textit{NuSTAR} & 60402021008 && 48 \\
                \hline
                \textit{XMM-Newton} & 0830551201 & 2019-02-02 & 35 \\
                \textit{NuSTAR} & 60402021010&  & 51 \\
                \hline
        \end{tabular}
\end{table}

\section{Timing properties}

\begin{figure*}
        \centering
        \includegraphics[width=0.98\textwidth]{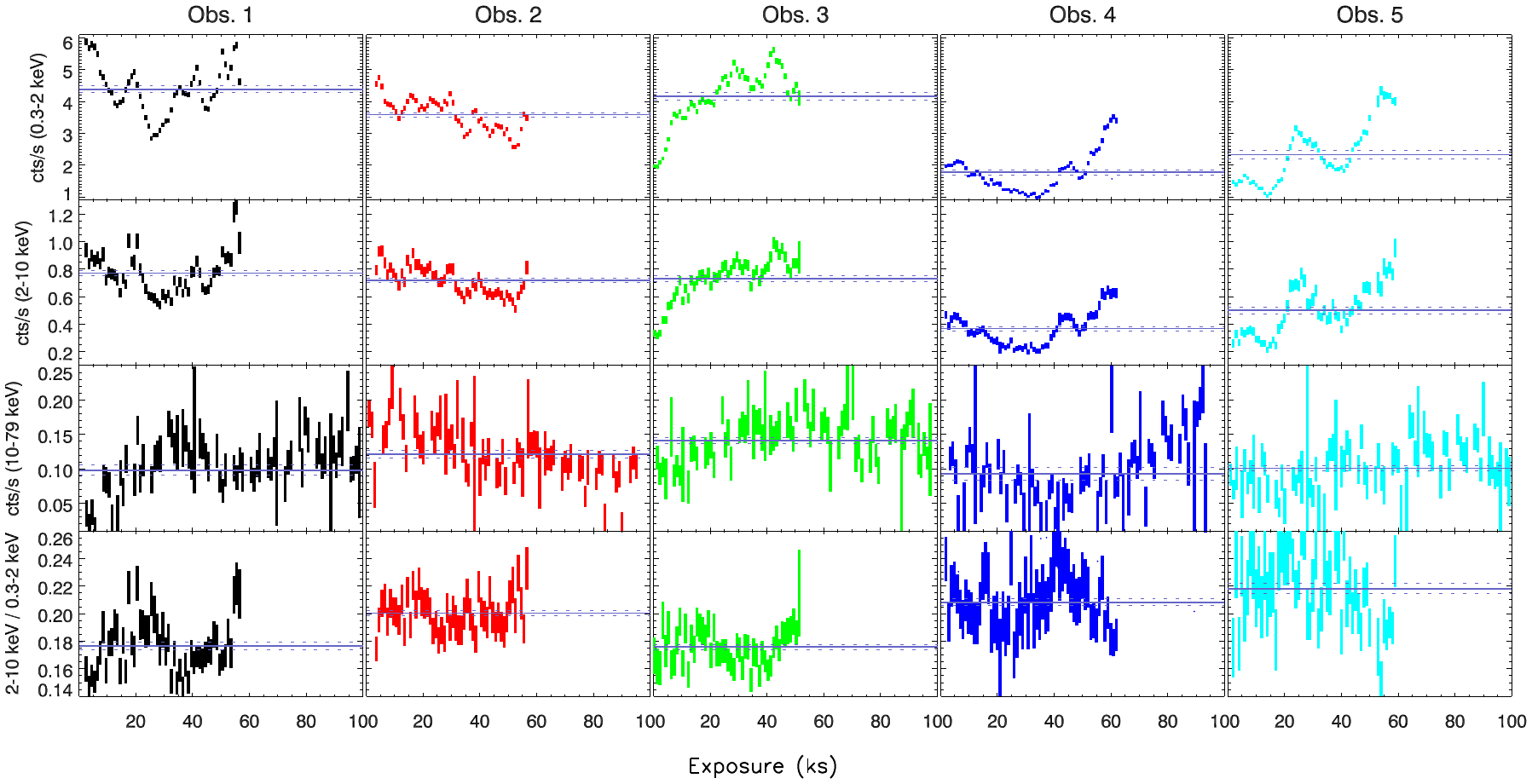}   
        \caption{\small{\textit{XMM-Newton}  light curves (background subtracted) in the 0.3-2 keV and 2-10 keV energy bands are shown in the top and top-middle panels, respectively. In the bottom-middle panels, the \textit{NuSTAR} background subtracted light curves in the 10-79 keV band are reported, while the bottom row reports on the ratios between light curves in the 0.3-2 keV and 2-10 keV band. The adopted time binning is 1 ks for all the observations, and the solid violet lines account for the average rate of each observation, while dashed lines account for the standard error of the mean.\label{lc}}}
\end{figure*}

\begin{figure}
                \centering
                \includegraphics[width=0.49\textwidth]{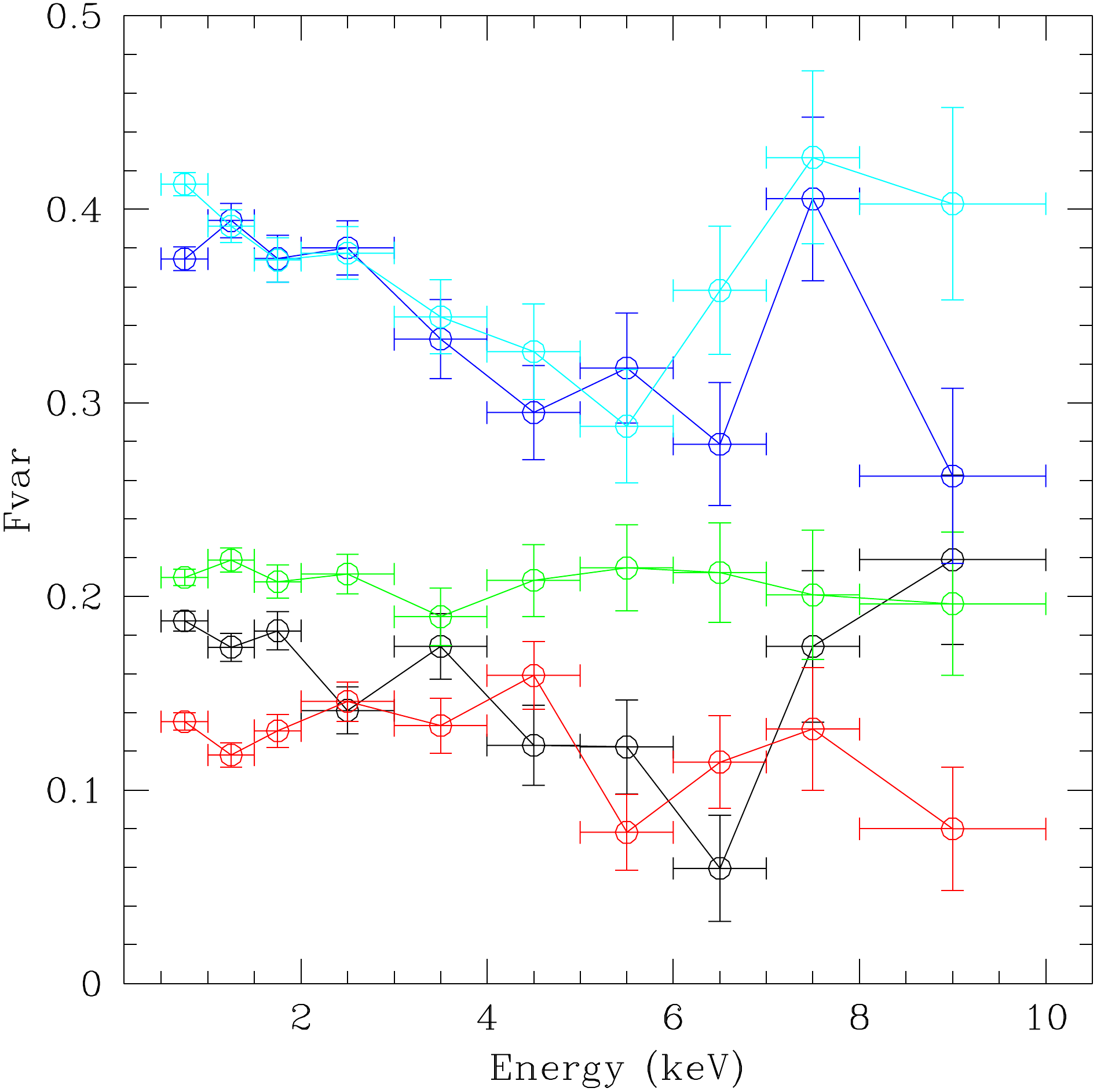}        
                \caption{\small{Fractional variability spectra for the different observations. Observations 1, 2, and 3 show smaller variability with respect to observations 4 and 5. Error bars account for 1$\sigma$ uncertainty}.
                \label{fvar}}
        \end{figure}

\indent The variability of an AGN is stochastic, aperiodic, and ubiquitously observed across the whole electromagnetic spectrum. In X-rays, AGNs not only exhibit long-term amplitude variations \citep[e.g.][]{Vagn11,Vagn16,Paolillo2017,Gallo18}, but also show rapid changes down to ks timescales \citep[e.g.][]{Ponti12}.\\
\indent During the monitoring, Mrk 359 varied both within and among the observations (see Fig.~\ref{lc}). In accordance with Fig.~\ref{lc}, we can define two different flux states in the \textit{XMM-Newton} bands: a high one in visits 1, 2, and 3, and a low one (counts decrease by a factor of $\sim$3) in the fourth and fifth observations. In Fig.~\ref{lc}, the two flux regimes are emphasised by the horizontal violet solid line, which indicates the average count rate of that specific observation in that band. Interestingly, UVM1 and UVW2 counts drop between observation 3 and 4 (see Fig.~\ref{om}).\\
\indent Flux variations are observed at timescales as short as a few ks both in the soft and hard X-ray bands. The largest amplitude variation occurs in observation 3, where the soft X-ray emission increases by a factor of $\sim$3 in under 50 ks. \textit{NuSTAR} light curves show a similar variability, though, in this high-energy band, and the presence of the two flux regimes seen in the \textit{XMM-Newton} data is not observed.\\
\indent In Fig.\ref{lc}, we also show the ratios of the (0.3-2 keV) and (2-10 keV) bands for the \textit{pn} data. The hardness ratios do not exhibit strong changes within each pointing, hence, we decided to use the time-averaged spectrum for the forthcoming spectral analysis. Ratios show that the source spectrum is hardening in observations 4 and 5 with respect to visits 1, 2, and 3 and this further supports the change of state during the campaign.\\
\indent As reported by several authors, the amount of X-ray variability at short timescales is anti-correlated with the AGN's luminosity \citep[e.g.][]{Barr1986,Green93,Lawrence93,Ponti12}, and the same behaviour has been found at longer timescales \citep[e.g.][]{Markowitz2001,Vagn11,Vagn16}. On the other hand, such an anti-correlation has been explained as a by product of a more intrinsic relation between the variability and SMBH mass itself \citep[e.g.][]{Papadakis2004,McHardy2006,Kording2007}. For these reasons,
 X-ray amplitude variations can be used to weight the SMBH mass. Following the prescriptions of \cite{Ponti12}, we computed the normalised excess variance \citep[e.g.][]{Nand97,Vaughan03}.
This estimator can be defined as $\sigma^2_{\rm{nxs}}= (S^2-\sigma_{\rm{n}}^2)/\mean{f}^2~$, where $S^2$ and $\sigma_{\rm{n}}$ account for the variance and mean square uncertainties of the fluxes, while $\mean{f}$ is the unweighted mean of the flux. Adopting a 250 s time bin for estimating the light curves and selecting 20 ks-long segments, we computed the normalised excess variance in the 2-10 keV band to be $\sigma^2_{\rm{nxs}}$=0.026$\pm$0.009. Then, following the relations presented by \citet{Ponti12}, we estimate the Mrk 359 BH mass to be M$_{\rm{BH}}$=(3.6$\pm$1.4)$\times10^6$ M$_\sun$.\\
\indent Such a variability-based estimate is consistent within the errors with the BH mass of M$_{\rm{BH}}$=1.7$\times10^6$ M$_\sun$ measured by \cite{Wang01}, who computed the stellar velocity dispersion $\sigma_\star$ via the [\ion{O}{III}].
Our estimate is also marginally compatible with what has been found by \citet{Williams2018} based on the X-ray scaling method \cite[firstly discussed in][]{Gliozzi2011}.
However, the above quoted measures are larger than the one by \citet{Botte05}, who obtained M$_{\rm{BH}}=6.9\times10^5$ M$_\sun$ via a [\ion{Ca}{II}] absorption triplet-based stellar velocity dispersion.\\
\indent We adopt our BH mass estimate to compute the bolometric luminosity and the Eddington ratio of Mrk 359. In particular, we relied on the bolometric correction described by \citet{Duras2020}. Such a bolometric correction was calibrated by fitting the spectral energy distribution of $\sim$1000 unobscured and obscured AGNs (type 1 and 2, respectively) of which luminosities cover six orders of magnitude. Following their paper and using an average 2-10 keV luminosity of 3$\times$10$^{42}$ erg s$^{-1}$, we obtain $\log$(L$_{\rm Bol}/ \rm erg~s^{-1})$=43.6 and an Eddington ratio of $\log$(L$_{\rm Bol}/L_{\rm Edd})\approx$8\%.
\\
\indent We further exploited the current data set by computing the fractional root    mean square variability amplitude (F$_{\rm var}$) for each observation of the monitoring \citep[e.g.][]{Edelson2002}. Such a variability estimator is defined as the square root of the normalised excess variance and it has been widely adopted in literature \citep[e.g.][]{Vaughan04,Ponti2006,Matzeu2016,Matzeu17,Alston2019,Parker2020,DeMarco2020,Igo2020}.    We compute F$_{\rm var}$ following \cite{Vaughan03} and using the background subtracted \textit{pn} light curves with a temporal bin of 2500 sec. The obtained F$_{\rm var}$ are shown in Fig.~\ref{fvar} as a function of the energy. These variability spectra further strengthen the idea of two different states: one less variable and corresponding to the higher flux state of observations 1, 2, and 3, and the other being characterised by larger variability and corresponding to the lower flux observations 4 and 5.
Interestingly, the decrease of variability with luminosity is naturally explained if the process results by the superposition of $N$ randomly emitting sub-units. Such a scenario, already considered for optical analyses \citep[e.g.][]{Pica1983,Aretxaga1997}, predicts a variability amplitude $\propto N^{-0.5}\propto L^{-0.5}$ when, in its simplified 0-th order version, the sub-units are identical and flare independently \citep[e.g.][]{Green93, Nand97,Almaini2000}. In the X-ray band, such a behaviour has been reported by many authors, both for local and high-redshift AGNs \citep[e.g.][]{Barr1986,Lawrence93,Papadakis2008,Vagn11,Vagn16}
In accordance with Fig.~\ref{fvar}, the different spectral bands vary in a similar amount, at least for the observations 1, 2, and 3, while in observations 4 and 5, the soft X-ray band may have larger variability.
As a final test, we searched for correlations between light curves at different energies or, in other words, we checked if the light curves at different energies varied in concert or not. In particular, we extracted the background subtracted light curves in the 0.5-2, 2.5-5 and 8-10 keV bands, these intervals likely being dominated by different emitting components, the soft excess, the primary continuum, and the reflected component. We therefore computed the Pearson cross-correlation coefficient for the different light curves, finding a strong correlation between the 0.5-2 and 2.5-5 keV bands (P$_{\rm cc}$=0.91, P(r<0)<10$^{-22}$) and a moderate correlation between the 2.5-5 and 8-10 keV bands  (P$_{\rm cc}$=0.74, P(r<0)<10$^{-15}$). Finally, the 0.5-2 keV light curve  and the 8-10 keV one are only weakly correlated, as supported by a P$_{\rm cc}$=0.42, P(r<0)=10$^{-6}$.

        .
\section{Spectral analysis}

In the fits, the Galactic hydrogen column density N$_{\rm{H}}$=4.38$\times$10$^{20}$ cm$^{-2}$ \citep{HI4PI} is always included and kept frozen to the quoted value. When \textit{OM} data are analysed, extinction is taken into account with the \textit{redden} model in \textit{XSPEC}. We used R$_{\rm{B-V}}$=0.0468 \citep{Schlafly11}, and we kept this value fixed during the computations.

\subsection{Primary emission and its reflected component} 
\begin{table*}
        \centering
        \setlength{\tabcolsep}{2.5pt}
        \caption{\small{Best fit parameters for the 3-10 keV \textit{XMM-Newton} data and for the 3-79 keV joint \textit{pn} and \textit{FPMA/B} data. Normalisations are in units of photons keV$^{-1}$ cm$^{-2}$ s$^{-1,}$ except for the line one, which is in photons cm$^{-2}$ s$^{-1}$. Observed fluxes are in units of 10$^{-12}$ erg cm$^{-2}$ s$^{-1}$. For each observation, we also quote the reflection fraction $R$ computed using \textit{xillver}.}
                \label{poxmm}}
        
        \begin{tabular}{c c c c c c c c c}
                \hline
                \setlength{\tabcolsep}{10.pt}
                Statistics&Band&Component & Parameter& Obs. 1& Obs. 2 & Obs. 3& Obs. 4& Obs. 5\\
                \\
                \hline
                \hline
        $\chi^2$/d.o.f.&3-10 keV &power law & $\Gamma$ &1.67$\pm$0.05 &1.71$\pm$0.08 &1.74$\pm$0.05&1.58$\pm$0.07&1.61$\pm$0.07\\
                443/448&&& Norm ($\times$10$^{-4}$)  &13.2$\pm$0.1 &12.4$\pm$0.1&14.5$\pm$0.1&5.7$\pm$0.6&8.6$\pm$0.1\\
                &&zgauss    &E (keV)  &6.40$\pm$0.03& 6.42$\pm$0.16 &6.39$\pm$0.12&6.37$\pm$0.06&6.37$\pm$0.07  \\
                &&&Norm  ($\times$10$^{6}$)   &4.7$\pm$1.8&5.5$\pm$3.7 &4.1$\pm$1.5&4.9$\pm$2.1&4.4$\pm$1.9\\
                &&&Eq.W (eV)    &80$\pm$35&80$\pm$60  &70$\pm$35&130$\pm$50&110$\pm$40 \\
                \hline

                $\chi^2$/d.o.f.&3-79 keV&       cut-off power law & $\Gamma$ &1.81$\pm$0.04&1.77$\pm$0.05&1.82$\pm$0.04&1.68$\pm$0.06&1.75$\pm$0.05\\
                1298/1242&&& E$_{\rm{c}}$ (keV) &>170&>140&>220&>135&>100\\
                &&& Norm  (10$^{-3}$) &1.6$\pm$0.1&1.5$\pm$0.1&1.6$\pm$0.1&0.6$\pm$0.5&1.0$\pm$0.1\\
                &&\textit{xillver}& A$_{\rm Fe}\dagger$&2.3$^{+0.8}_{-1.1}$&&&&\\
                &&&$R$&0.40$\pm$0.10&0.26$\pm$0.09&0.41$\pm$0.10&0.50$\pm$0.15&0.44$\pm$0.13\\
                &&& Norm  (10$^{-5}$) &1.7$\pm$0.4&1.2$\pm$0.4&1.7$\pm$0.4&1.4$\pm$0.3&1.4$\pm$0.5\\
                &&&F$_{3-10~\rm keV}$   &4.4$\pm$0.1&4.4$\pm$0.1&4.3$\pm$0.1&2.3$\pm$0.1&3.2$\pm$0.1\\
                &&&F$_{10-78~\rm keV}$
                &13$\pm$1&11$\pm$2&12$\pm$1&8$\pm$3&10$\pm$2\\
                \hline
                \hline
        \end{tabular}
\end{table*}
\begin{figure}
        \centering
        \includegraphics[width=0.49\textwidth]{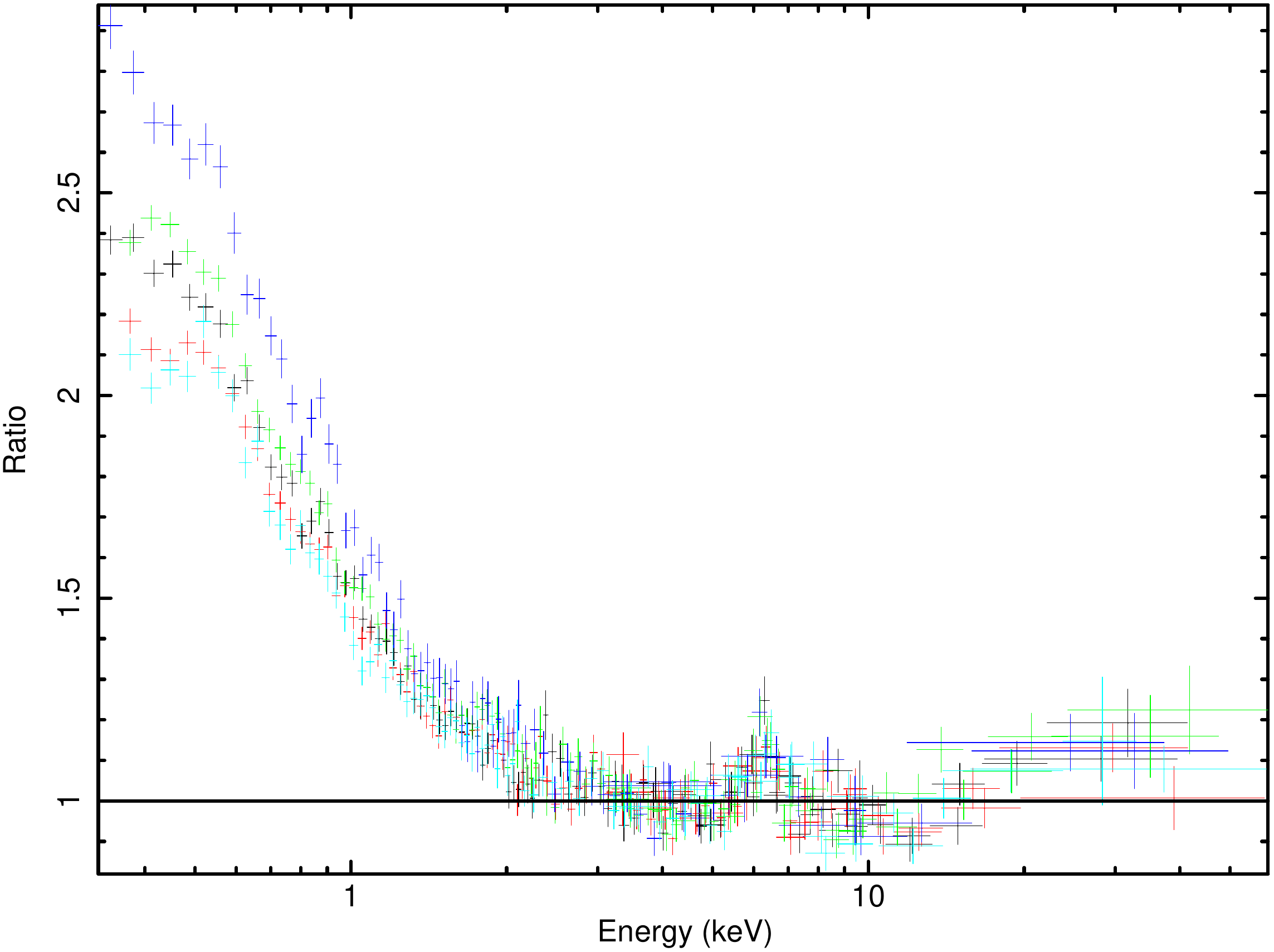}               
        \caption{\small{Joint \textit{XMM-Newton-NuSTAR} spectra with respect to a power-law modelling in the 3-10 keV band. A remarkable soft excess rises smoothly below 2-3 keV above the extrapolated 3–10 keV power law in all the observations. Moreover, it exhibits strong variability. In the Fe K energy band, a strong emission line is apparent, while in hard X-rays, residuals show a Compton hump. \label{soft_excess}}}
\end{figure}

\begin{figure}
        \centering      
        \includegraphics[width=0.49\textwidth]{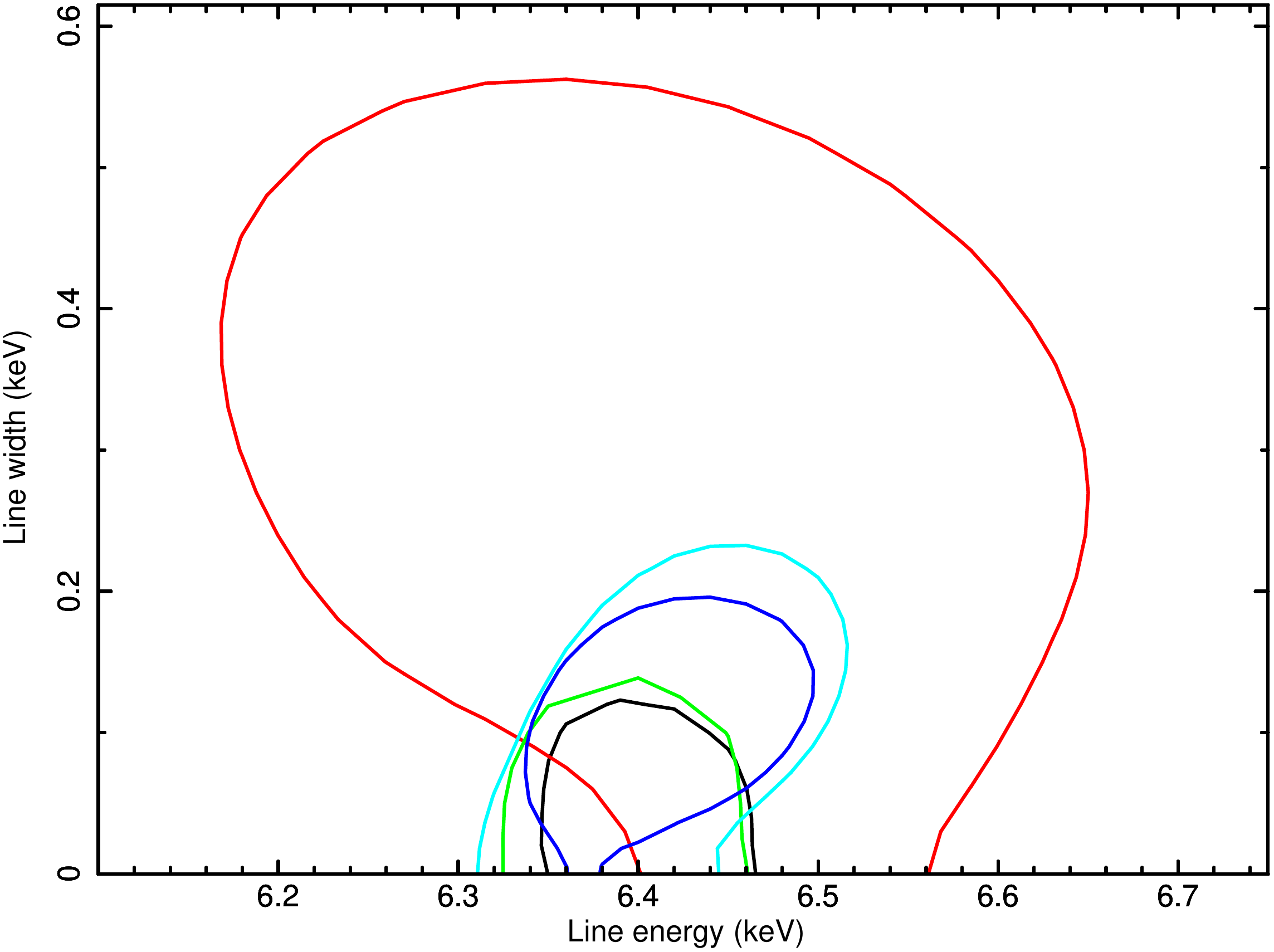}            
        \caption{\small{ Confidence contours at 90\% (two significant parameters) confidence levels for the Fe K$\alpha$ energy centroid and line intrinsic width. The Fe K$\alpha$ width is consistent with zero in all the observations \label{width}}.}
\end{figure}

We started to focus on the Fe K$\alpha$ properties only using \textit{pn} data, and ignoring spectra below 3 keV. We adopted a power law to fit the 3-10 keV spectra, with the model parameters untied between the observations and free to vary. This crude fit, only accounting for the primary continuum, returns a $\chi^2$=525 for 458 degrees of freedom (d.o.f.). This is mainly due to the prominent emission line at about 6.4 keV, see Fig.~\ref{soft_excess}. To model such residuals, we add a Gaussian component. We therefore tested the model: \textit{tbabs$\times$(power-law+zgauss)} in which the line energy centroid, its width, and normalisation were free to vary in all the observations. However, this preliminary fit shows that in none of the pointings is the line resolved (see Fig.\ref{width}), thus for the forthcoming fits, we fixed the line width to zero.
A value of $\chi^2$=443 for 448 d.o.f characterises our model and the corresponding best fit parameters are reported in Table \ref{poxmm}. The Fe K$\alpha$ fluorescent emission-line intensity is consistent with being constant over the campaign. This, together with the narrow profile, suggests an origin from material far from the nuclear region.

To constrain and understand the nature of the reprocessed emission in Mrk 359, we added the \textit{NuSTAR FPMA/B} 3-79 keV data. \textit{\emph{The }EPIC-pn} spectra showed the Fe K$\alpha$ to be constant and narrow, likely originating from Compton-thick cold material far from the SMBH. Therefore, we studied the 3-79 keV spectra using a cut-off power law to account for the source's primary continuum and \textit{xillver} \citep[][]{Garc14a,Daus16} to reproduce the neutral Fe K$\alpha$ line and its associated reflection spectrum. In the fit, we left the high-energy cut off and the primary normalisation untied and free to vary between the observations
the photon index. \textit{Xillver}'s continuum parameters are tied to those of the cut-off power law, with the exception of the normalisation, which is set as free to vary in each observation.
The Fe abundance is fitted by tying its value among the pointings, while the ionisation parameter is fixed to $\log(\xi/\rm erg~s^{-1} cm)$=1 (close to neutral). To account for the inter-calibration between the detectors, we include a free constant. The two \textit{NuSTAR} modules are in good agreement with each other ($\sim$3\%), and they agree with \textit{pn} data within $\lesssim$7\% in all but the fourth observation, for which a value of about 30\% is found. Such a high value is due to the flux increase caught by \textit{NuSTAR} only (see the blue light curves in Fig.~\ref{lc}). We re-extracted the \textit{NuSTAR} spectra in order to be truly simultaneous with \textit{XMM-Newton}. Once truly simultaneous \textit{XMM-Newton/NuSTAR} data are considered, the different instruments are consistent with each other within $\sim5\%$, thus, in the subsequent analysis, we used these truly simultaneous spectra.
\begin{figure}
        \centering      
        \includegraphics[width=0.48\textwidth]{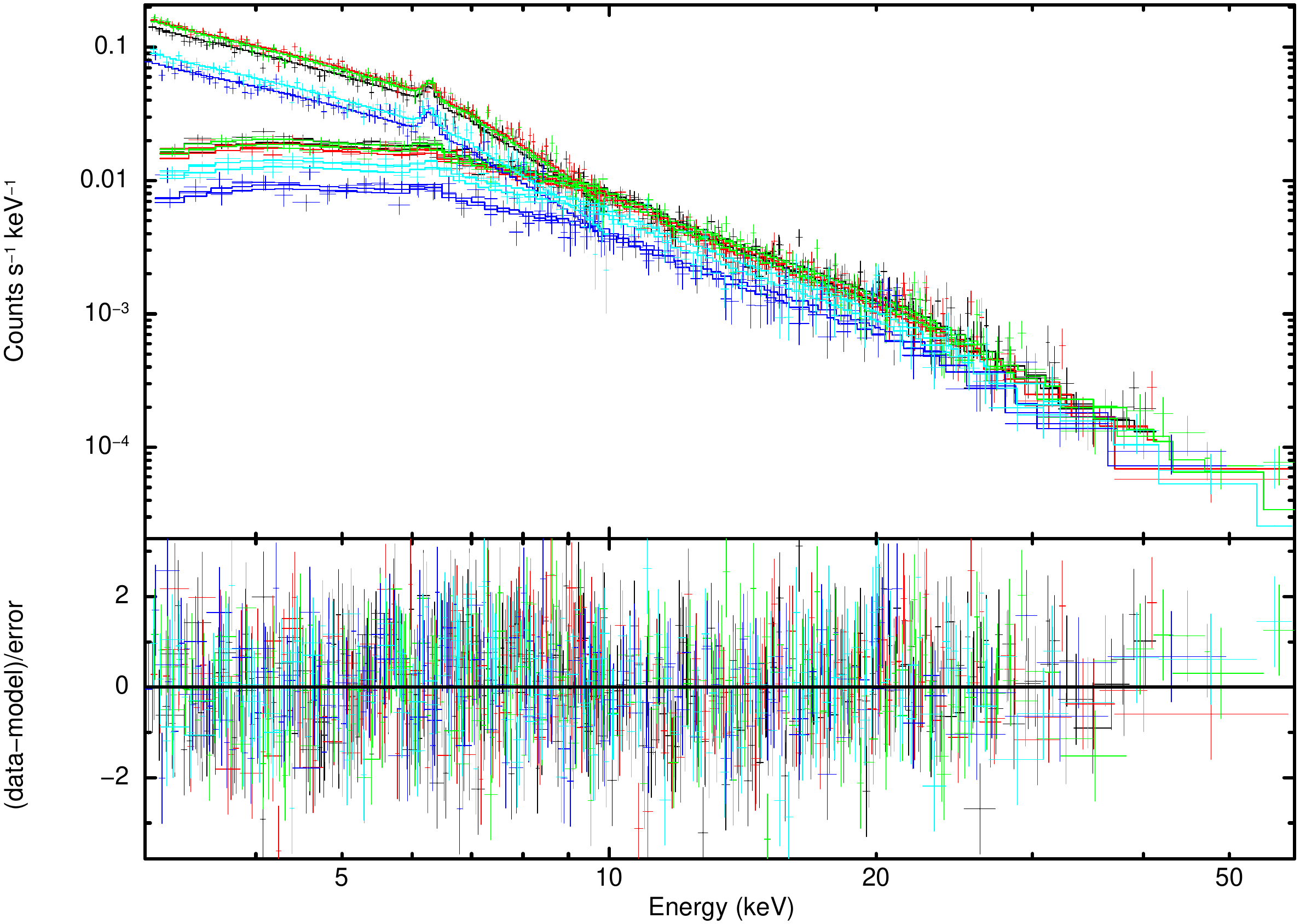}       
        \caption{\small{Best fit to the 3-79 keV \textit{XMM-Newton-NuSTAR} data corresponding to the model (\textit{tbabs$\times$cutoffpl$\times$xillver}). $\chi^2/{\rm d.o.f.}$=1.04. \label{379xmm_nustar}}}
\end{figure}
\begin{figure*}
        \centering
        \includegraphics[width=0.49\textwidth]{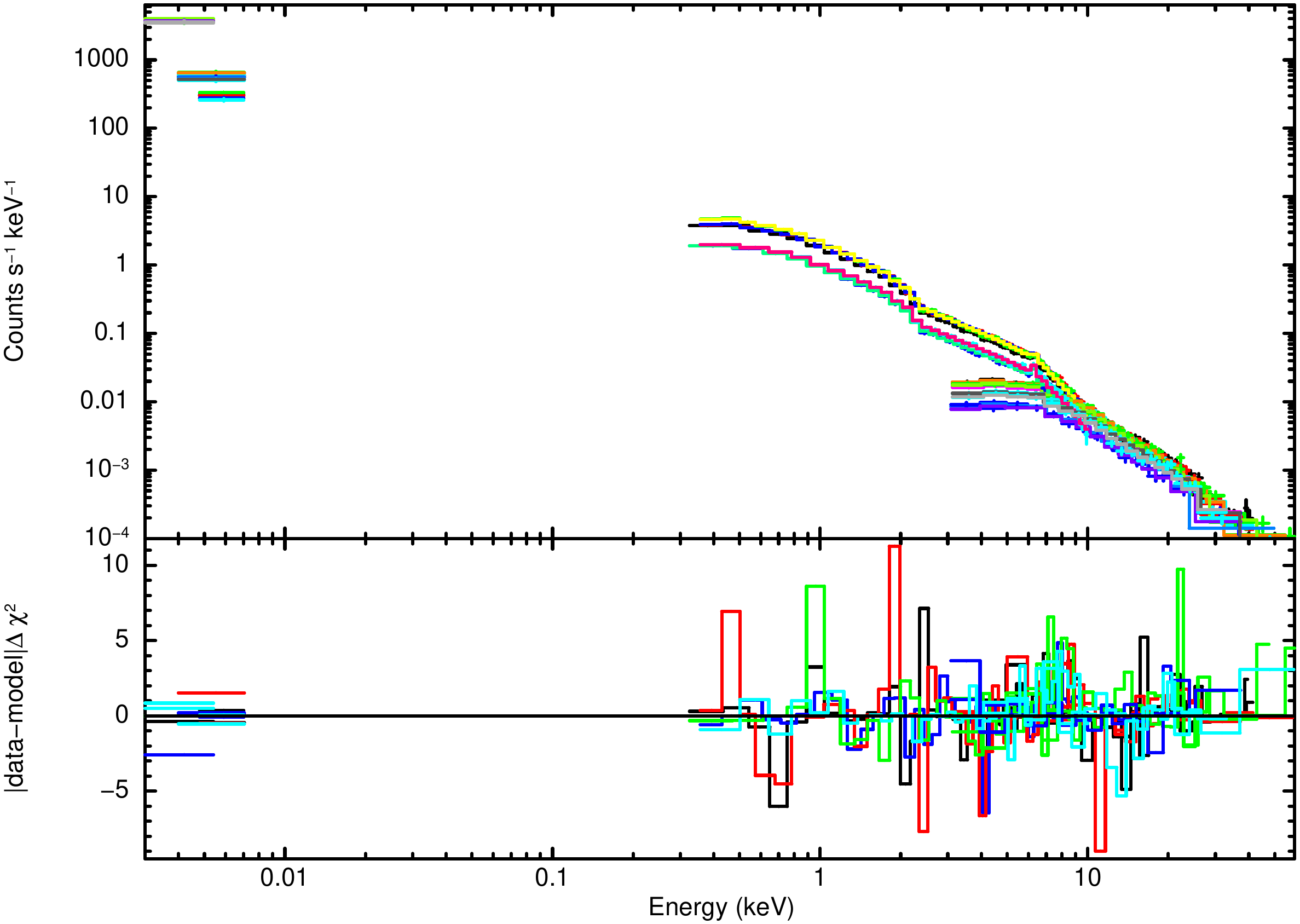}
        \includegraphics[width=0.465\textwidth]{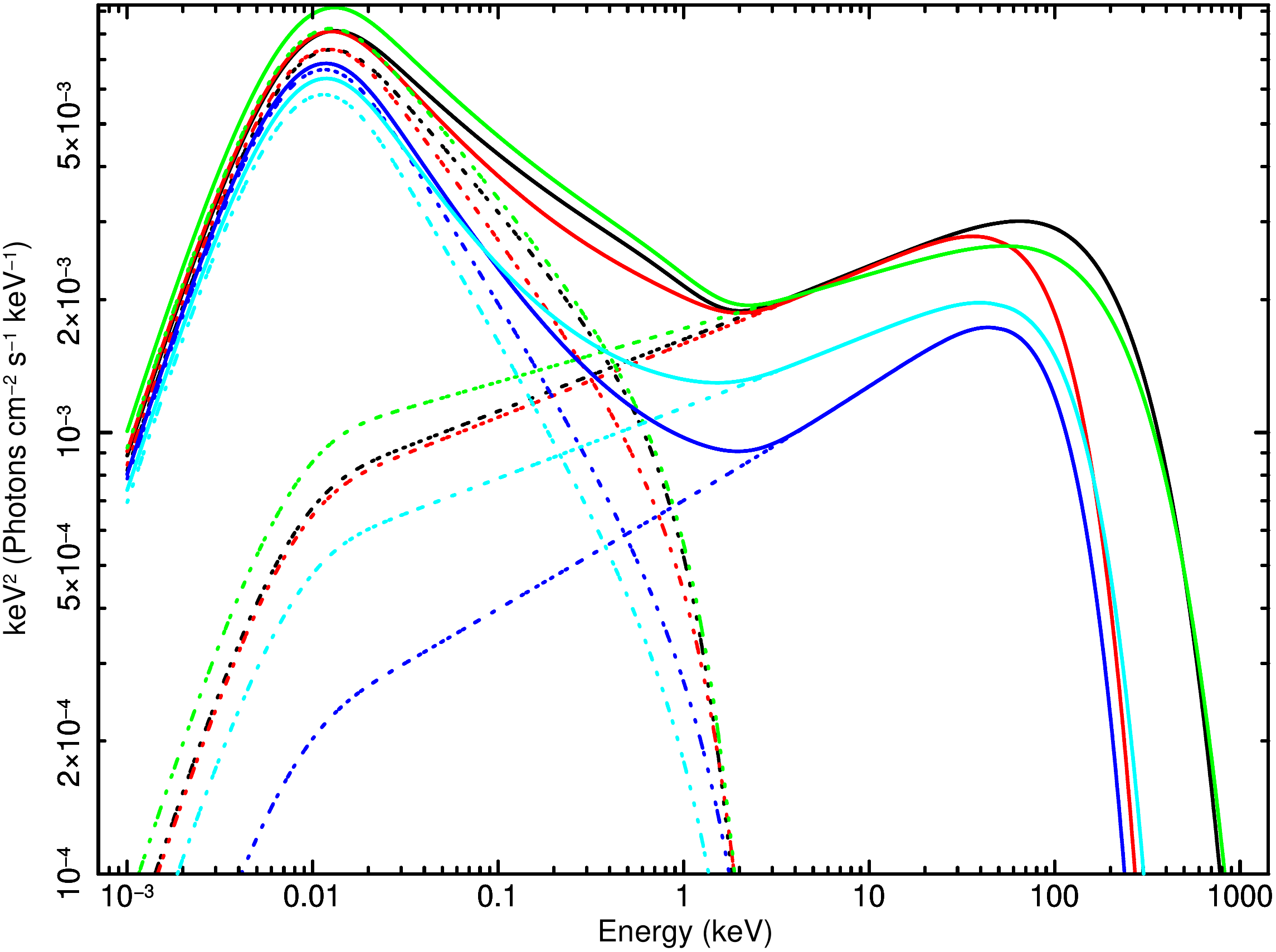}                            
        \caption{Left panel: optical-UV to hard X-rays fitted with the two-corona model ($\chi^2=$ 1690 for 1567 d.o.f.). Right panel: underlying best fit model is shown by the solid lines, while dashed ones account for the contribution of each of the two coronae.\small{\label{besttwocorona}}}
\end{figure*}   
These steps lead to a best fit with $\chi^2$=1298 for 1242 d.o.f.. In Table~\ref{poxmm}, we report the corresponding best fit values. We notice that this model yields higher values for the photon index of the nuclear emission now being in the range 1.68-1.82. Though the photon indices are fairly consistent within the errors, the overall scenario is consistent with the softer when brighter behaviour, in agreement with what commonly observed in nearby Seyfert galaxies \citep[e.g.][]{Sobolewska09} or in AGNs samples \citep[e.g.][]{Serafinelli17}. The high-energy cut off (E$_{\rm c}$) is unconstrained, and the higher value for its lower limits is E$_{\rm c}$>220 keV, found in observation 3. To further assess the constancy of the reflected emission in Mrk 359, we tied the \textit{xillver} normalisation among the observations. Moreover, we also try to fit the high-energy cut off  tying its value between the pointings. This procedure returns a fit of $\chi^2$=1302 for 1250. Only a lower limit is found for the high-energy cut off  (E$_{\rm c}$> 340 keV). The model \textit{xillver} normalisation is N$_{\rm refl}$=(1.5$\pm$0.3)$\times$10$^{-5}$ photons keV$^{-1}$ cm$^{-2}$ s$^{-1}$ with an associated flux F$_{\rm 3-79 ~keV}$= (8.7$\pm$1.4)$\times$10$^{-13}$ erg s$^{-1}$ cm$^{-2}$.

\subsection{Broadband data: testing the two-corona model}

\begin{figure}
        \centering      
        \includegraphics[width=0.37\textwidth]{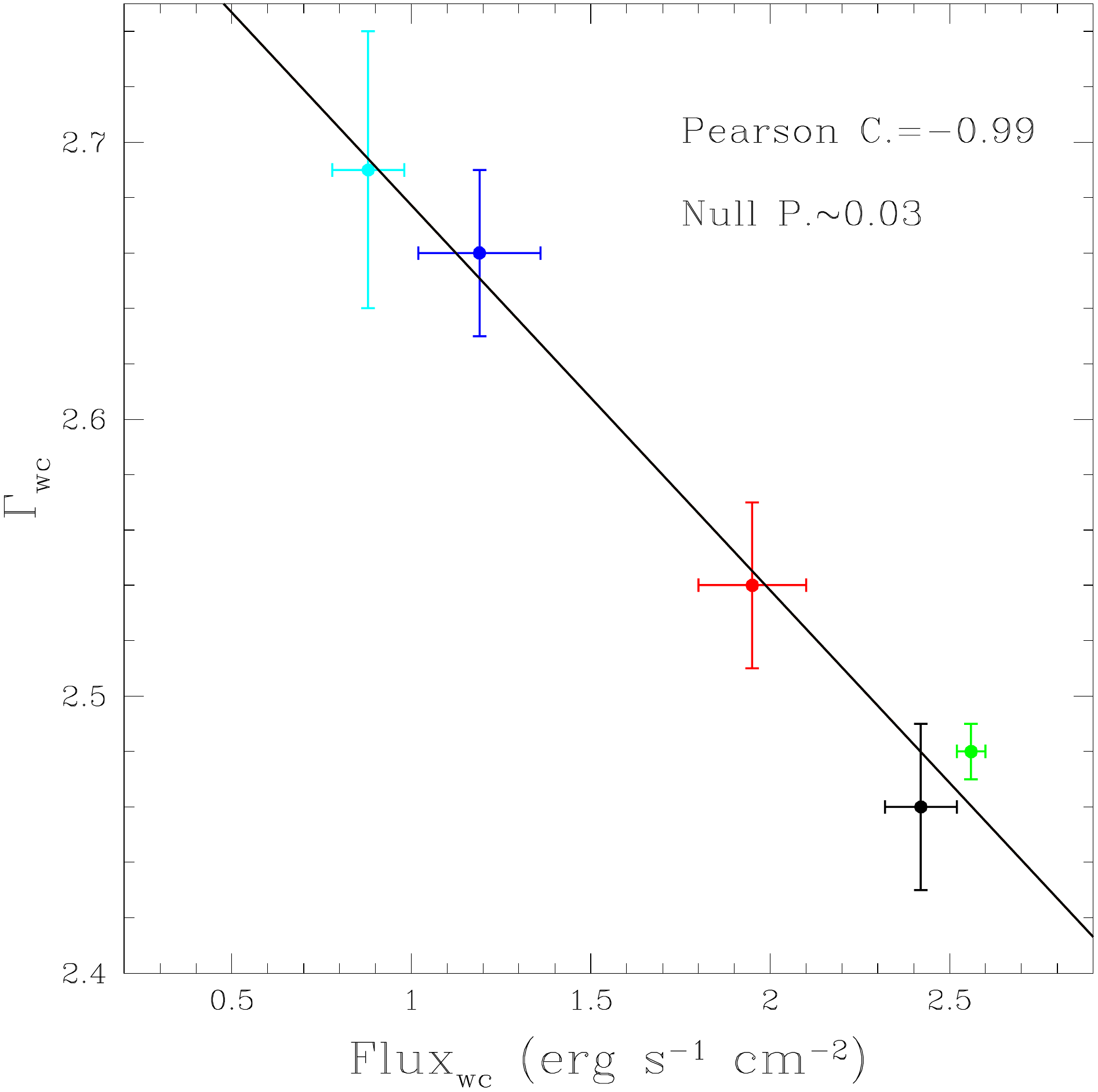}    
        \includegraphics[width=0.37\textwidth]{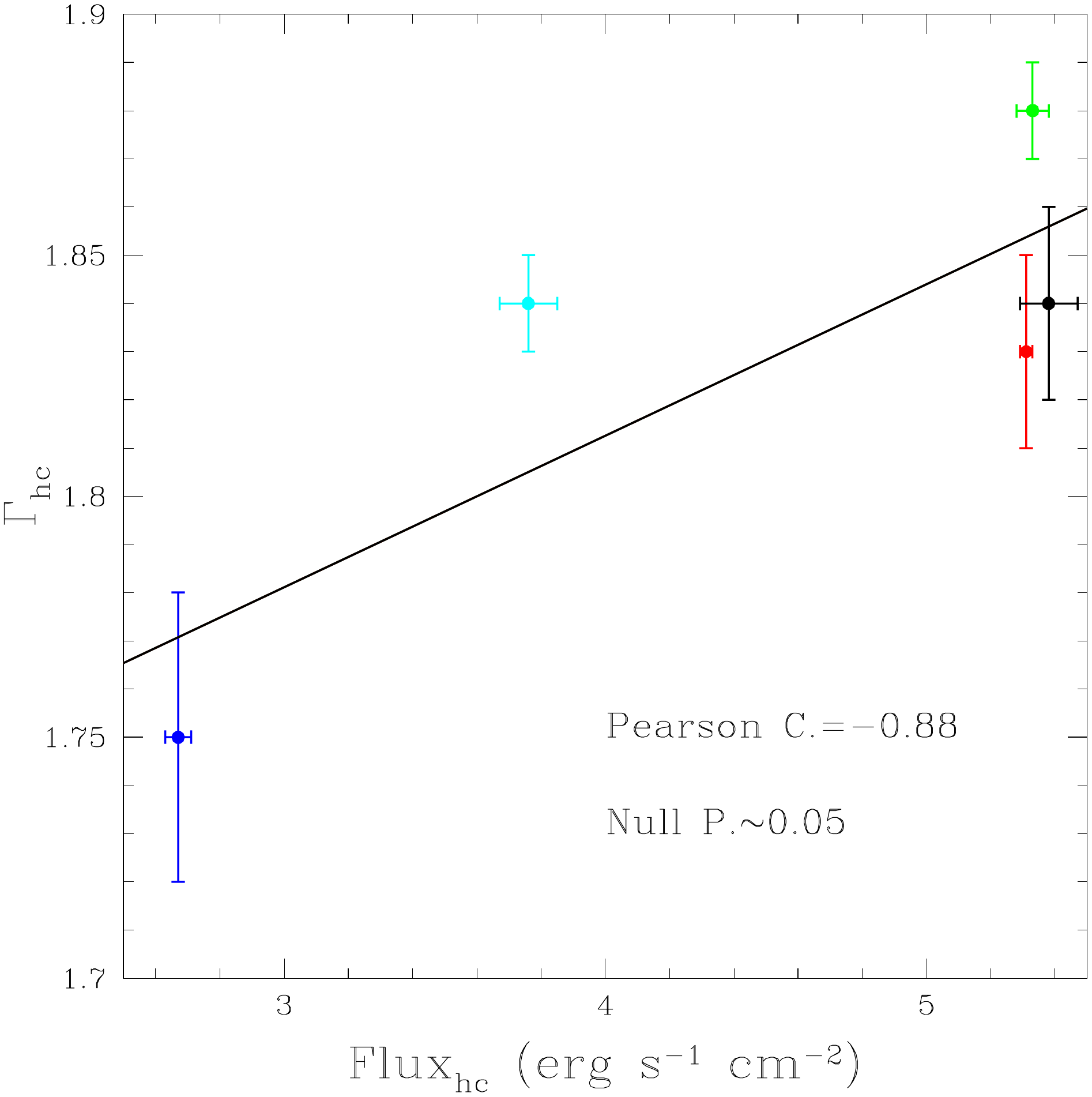}     
        \caption{\small{ Top panel: $\Gamma_{\rm wc}$ as a function of the warm Comptonisation flux estimated in the 0.3-2 keV. Bottom panel: the same as for the hot corona, for which the flux has been computed in the 2-10 keV band. Fluxes are in units of 10$^{-12}$ erg s$^{-1}$ cm$^{-2}$ \label{gamsh}}}
\end{figure}

\begin{table*}
        \centering
        \setlength{\tabcolsep}{2.5pt}
        \caption{\small{Best fit parameters for the two-corona model. Normalisations are in units of photons keV$^{-1}$ cm$^{-2}$ s$^{-1}$. A $\dagger$ is used for the parameters fitted but tied between the observations, while the corona opacities are derived from the best fit values.}
                \label{TC}}
        \begin{tabular}{c c c c c c c c}
                \hline
                \setlength{\tabcolsep}{10.pt}
                                \\
                $\Delta\chi^2$/dof&Component & Parameter& Obs. 1& Obs. 2 & Obs. 3& Obs. 4& Obs. 5\\

                \hline
                \hline
                
                1690/1567&nthcomp$_{\rm wc}$ & $\Gamma_{\rm{wc}}$ & 2.46$\pm$0.03&2.54$\pm$0.03&2.48$\pm$0.01&2.66$\pm$0.03&2.69$\pm$0.05\\
                && kT$_{\rm{wc}}$  (keV) &0.26$\pm$0.02 &0.31$\pm$0.03&0.27$\pm$0.01&0.38$\pm$0.05&0.32$\pm$0.07\\
                && $\tau_{\rm{wc}}$ &28.2$\pm$1.1&25.1$\pm$1.5&27.8$\pm$0.7&21.4$\pm$1.7&23.1$\pm$2.1\\
                && kT$_{\rm{bb}}\dag$ (eV) &3.2$\pm$0.5&&&&\\
                &&Norm ($\times$10$^{-4}$) &5.2$\pm$0.5 &4.2$\pm$0.6&5.4$\pm$0.2&2.7$\pm$0.4&1.7$\pm$0.5\\
                \hline
                &nthcomp$_{\rm hc}$ & $\Gamma_{\rm{hc}}$ &1.84$\pm$0.02 &1.83$\pm$0.02&1.88$\pm$0.01&1.75$\pm$0.03&1.84$\pm$0.01\\
                && kT$_{\rm{hc}}$   (keV) &>35&>15&>55 &>25&>23\\
                && $\tau_{\rm{hc}}$ &<2.5&<4.3&<1.7&<3.1&<3.3\\
                &&Norm   ($\times$10$^{-3}$)&1.63$\pm$0.02 &1.58$\pm$0.05&1.72$\pm$0.04&0.69$\pm$0.05&1.14$\pm$0.04\\
                \hline
                &xillvercp&A$_{\rm{Fe}}\dag$  &1.3$\pm$0.3 &&&&\\
                &&$\log \xi\dag$   &1.33$\pm$0.05 &&&&\\
                &&Norm ($\times$10$^{-5}$)  &1.3$\pm$0.2 &1.2$\pm$0.2&1.7$\pm$0.2&0.9$\pm$0.2&1.5$\pm$0.3\\
        \end{tabular}
\end{table*}
\begin{figure}
        \centering
        \includegraphics[width=0.48\textwidth]{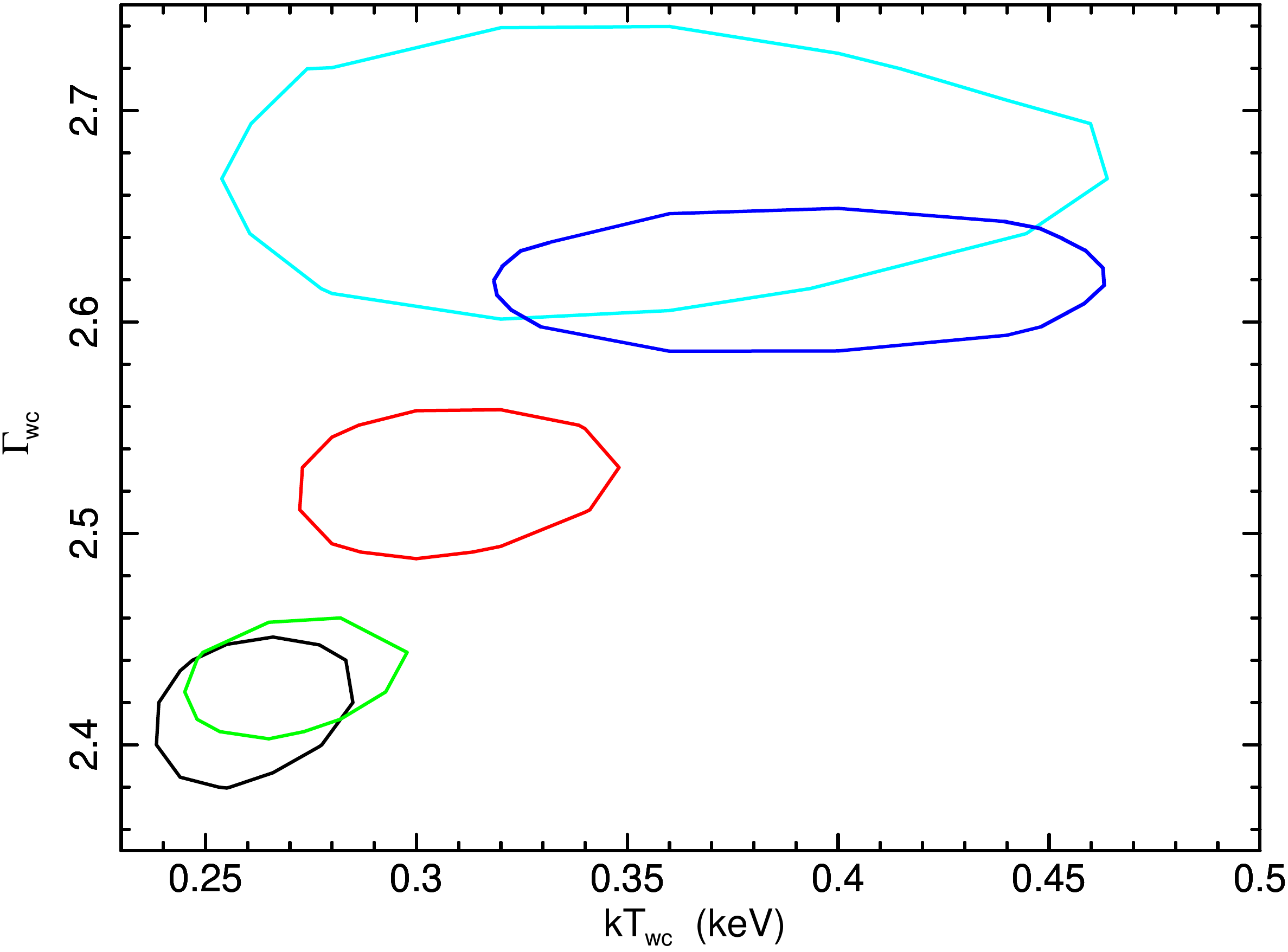}      
        \includegraphics[width=0.48\textwidth]{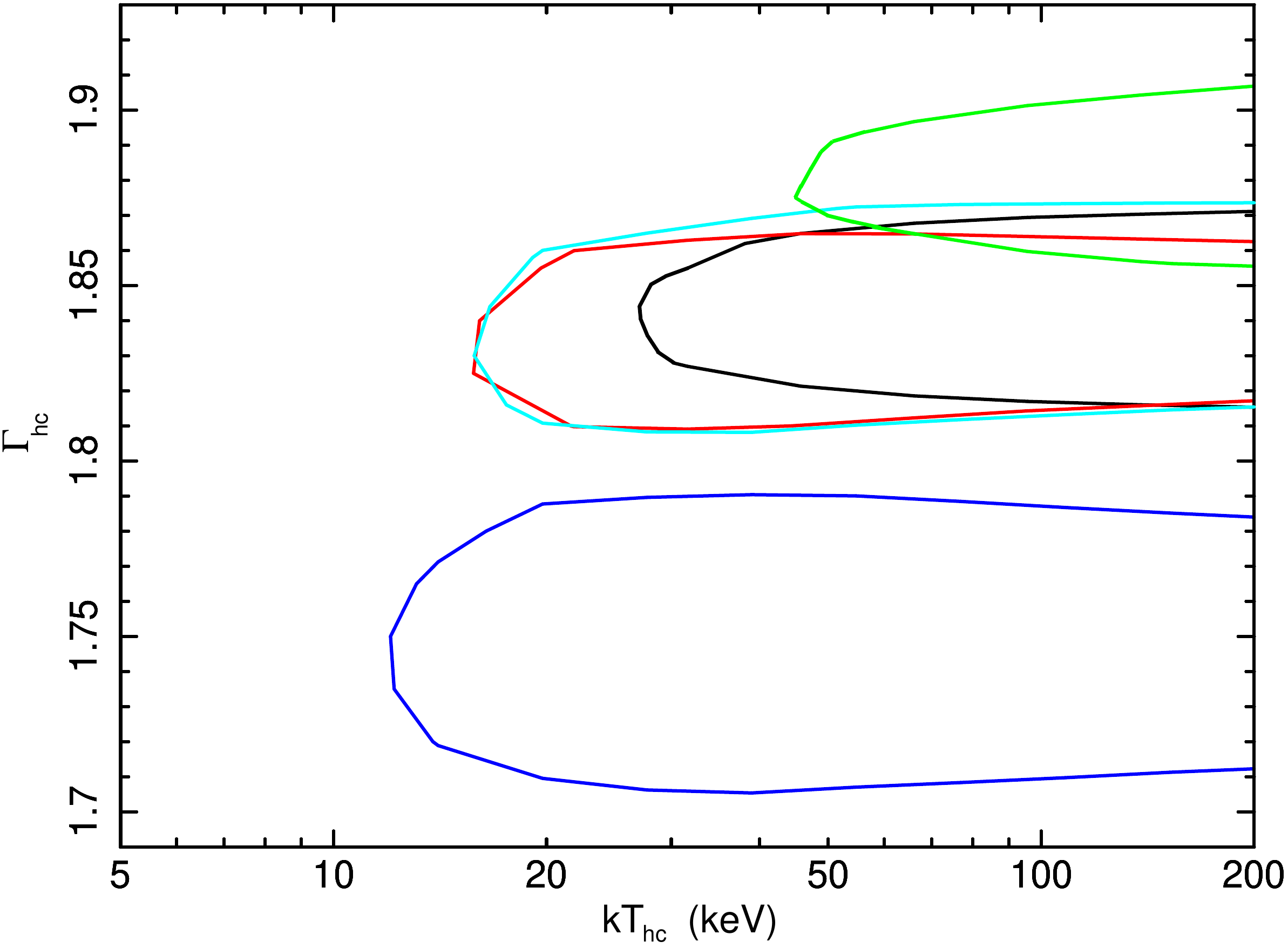}             
        \caption{\small{Contours at 90\% confidence level ($\Delta\chi^2$=4.61) between the primary continuum photon index and the temperature for both the warm corona (top panel) and the hot corona (bottom panel). The temperature of the hot corona remains unconstrained, and only lower limits are obtained. The photon indices of both components change from one observation to the other.\label{cont}}}
\end{figure}
At this stage of the analysis, we include \textit{OM} spectral points and \textit{pn} data down to 0.3 keV.
Two Comptonising components, one optically thick and warm (warm corona, wc) and the other optically thin and hot (hot corona, hc), have often been used to reproduce broadband AGN spectra \citep[e.g.][]{Porquet2018,Middei18,Ursini18,Noda2018,Kubota2018,Petrucci18,Middei19,Ursini2019}, therefore we tested the so-called two-corona model on the present dataset. In \textit{Xspec}, such a model appears as:\\
\\
 \textit{redden}$\times$~\textit{TBabs}$\times$~\textit{const}$\times$~\\
$\times$[small$_{\rm BB}$+~\textit{nthcomp$_{\rm{wc}}$}+\textit{nthcomp$_{\rm{hc}}$}+\textit{xillvercp}].\\
\\
The small$_{\rm BB}$ component accounts for the broad-line region (BLR), and in particular for the so-called small blue bump (SBB) at about 3000 $\AA$, as described in detail in \citet{Mehdipour2015}. By fitting this component (kept tied among the pointings), we find its flux to be F=4.5$\times$10$^{-12}$ erg cm$^{2}$ s$^{-1}$. 
\textit{Nthcomp$_{\rm{wc}}$} \citep{Zdzi96,Zyck99} provides a thermally Comptonised continuum whose high- and low-energy rollovers are parametrised by the temperatures of the electrons (kT$_{\rm wc}$) and the seed photons (kT$_{\rm bb}$). We assumed the seed photons to arise from a disc blackbody, and we fitted the kT$_{\rm{bb}}$ parameter by tying its value between the various observations. The photon index ($\Gamma_{\rm wc}$), the warm corona's temperature, and the normalisation are considered as free parameters in all the observations. Then, \textit{nthcomp$_{\rm hc}$} and \textit{xillvercp} are used to reproduce the high-energy emission of Mrk 359. The first model accounts for the primary Comptonised power law, the latter for its associated reflected component. Disc photons accounted by \textit{nthcomp$_{\rm hc}$} are assumed to arise from the same disc-like blackbody feeding the warm corona, thus we tied the kT$_{\rm bb}$ to the one in \textit{nthcomp$_{\rm wc}$}. To model the hot component, we fitted the primary photon index ($\Gamma_{\rm hc}$), the electrons temperature kT$_{\rm hc,}$ and normalisation in all the observations. The normalisation of \textit{xillvercp} is free for all the visits of the campaign, while the ionisation degree and the iron abundance are free to vary but tied among the different observations.\\
\indent Such a procedure yields the data best fit ($\chi^2$=1690 for 1567 d.o.f.) shown in the left panel of Fig.~\ref{besttwocorona}. The contribution of the two coronae to the fit is shown in the right panel of  Fig.~\ref{besttwocorona}. The best fit values for the parameters are reported in Table \ref{TC}. The fit returns an iron abundance A$_{\rm{Fe}}$=1.3$\pm$0.3 and an ionisation parameter of $\log(\xi/\rm erg~s^{-1}  cm)$=1.33$\pm$0.05. Using the best fit values for the coronal temperature and the photon index, we derived the opacity of the two coronae\footnote{In \textit{nthcomp,} the medium opacity is related to the photon index $\Gamma$ of the asymptotic power law and to the temperature $\theta$=kT$_{\rm e}$/m$_{\rm e}$ c$^{2}$ by the relation $\tau$={2.25+3/[$\theta\times$($\Gamma$+0.5)$^2$-2.25]}$^{1/2}$-1.5, where the subscript 'e' refers to the electron.}. The photon index of the warm component varies in the range 2.46-2.69 and higher values of $\Gamma_{\rm{wc}}$ are measured when the component flux decreases, see top panel of Fig.~\ref{gamsh}. Concerning the hot corona, the photon index shows weak changes (1.75$\leq\Gamma_{\rm hc}\leq$1.88) and only lower limits are found for the hot corona's temperature. Opposite to what we found for the warm corona, steeper $\Gamma_{\rm hc}$ correspond to brighter states.
For both coronae, we show the contours between the photon index $\Gamma_{\rm wc/hc}$ and the electron temperature kT$_{\rm wc/hc}$ in Fig.~\ref{cont}.

\subsection{Broadband data: testing relativistic reflection}

\indent A weak or negligible broad component of the Fe K$\alpha$ emission line may still allow for a prominent relativistic reflection continuum. For this reason, we investigated whether a blurred ionised reflection could be responsible for the soft excess in Mrk 359. Here, we modify the two-corona model presented in Sect. 4.2, replacing \textit{nthcomp$_{\rm wc}$} with \textit{relxillcp} \citep{Dauser2014,Daus16}. Such a model calculates a standard relativistic reflection spectrum resulting from a Comptonised continuum irradiating the accretion disc. Moreover, we account for the accretion disc emission including a \textit{diskbb} component. Therefore, we use the following model: 
\\

\noindent\textit{redden}$\times$~\textit{TBabs}$\times$~\textit{const}$\times$\\
$\times$~[small$_{\rm BB}$+\textit{diskbb}+\textit{nthcomp}+\textit{relxillcp}+\textit{xillvercp}].\\

\begin{figure}
        \centering      
        \includegraphics[width=0.48\textwidth]{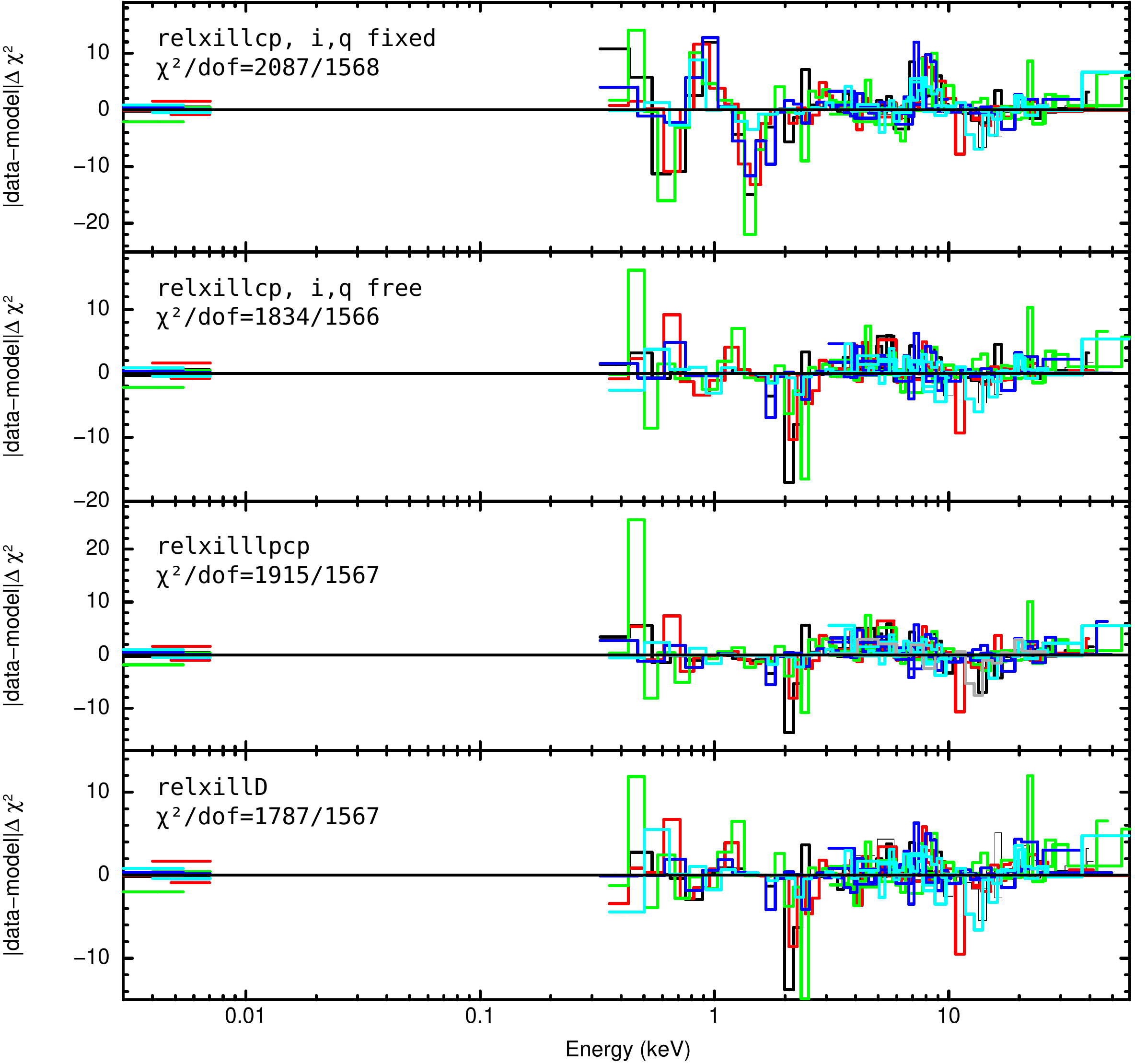}    
        \caption{\small{Contribution to the $\chi^2$ in each bin for the different relativistic models tested in Sect. 4.3. We recall the corresponding model in each panel's row, and for the first and the second rows we also specify some of the frozen or free parameters \label{refl}}.}
\end{figure}

\textit{Nthcomp} still provides the primary continuum emission, and we calculated the photon index, the electron temperature, and the normalisation in all the pointings. The \textit{nthcomp} seed photons temperature is the same as \textit{diskbb}, thus we tied this parameter between the two components. The current dataset does not provide any constrain on this temperature, and we set its value to be consistent with what was previously found for the two-corona model. The normalisation of the \textit{diskbb} component is free to vary in all observations. 
We use the \textit{small$_{\rm BB}$} table to account for the SBB also in the current model. During the fitting procedure, the table's normalisation was tied between the observations. After fitting the whole dataset, for this component we obtain a flux of $\sim$4$\times$10$^{-12}$ erg cm$^{-2}$ s$^{-1}$.
\textit{Relxillcp} and \textit{xillvercp} were used to reproduce the reflected continuum only, hence we fitted the models' normalisations in each pointing. \textit{Relxillcp} allows for spin ($a$) estimates, though after a preliminary fit we find this parameter to be consistent with $a$>0.985. For this reason, we assumed the SMBH to be maximally rotating and we left the inner radius free to vary in all the pointings. The iron abundance of the reflecting material is free to vary and tied between the reflection models and the observations. We also fitted the ionisation degree of the relativistic reflection component and that of \textit{xillvercp}.
Such a procedure returns a fit with a value of $\chi^2$/dof=2087/1568, with large residuals mainly in the soft X-ray band (see top row in Fig.~\ref{refl}). As a subsequent step, we allowed the corona emissivity $q$ and the inclination parameter $i$ to vary. The inclination is tied between \textit{xillvercp} and \textit{relxillcp}. The addition of these two parameters is beneficial in terms of statistical quality ($\Delta\chi^2/\Delta$dof=-253/-2), and we get an overall fit characterised by $\chi^2$/d.o.f.=1834/1566. In Table~\ref{relativistic}, the fitted values are shown. The current model returns a slightly supersolar iron abundance A$_{\rm Fe}$=1.55$\pm$0.25, and a $\log(\xi/{\rm erg~s^{-1}~cm})$=2.72$\pm$0.04 is required to reproduce the source soft-excess. The model returns a disc inclination of $i$=71$\pm$1 deg and the inner disc region is consistent with a few gravitational radii. Reflection due to distant matter is compatible with being constant and likely due to almost neutral material $\log(\xi/{\rm erg~s^{-1}~cm})=0.15\pm$0.12. The primary continuum can be described by a power law of which $\Gamma$ varies between 1.92 and 2.03, while the electron temperature is kT$\textgreater$140 keV. We notice that these photon indices are steeper and not consistent with those obtained in the two-corona model. \textit{Relxillcp} also allows for a broken power-law emissivity profile of the primary continuum emission ($q_2$). However, the assumption of such a different emissivity profile has a negligible impact in terms of statistical quality and $q_2$>1.3 is found.
\begin{table*}
        \centering
        \setlength{\tabcolsep}{5 pt}
        \caption{\small{Fitted values for the models tested in Sect. 4.3. Normalisations are in units of photons keV$^{-1}$ cm$^{-2}$ s$^{-1}$.}}\label{relativistic}
        \begin{tabular}{c c c c c c c c c c }
                \hline
                \\
                \textit{Relxill} version&$\chi^2$/dof&  Component&Parameter &Obs. 1& Obs 2. &Obs. 3& Obs. 4& Obs. 5\\
                \hline
                \hline
                standard        &1834/1566&diskbb&      T$_{\rm bb}$ ($\times10^{-3}$ eV)&3.4$\pm$0.5&&&& \\
                &&&Norm ($\times10^{9})$ &6.5$\pm$0.3&6.8$\pm$0.3&7.5$\pm$0.4&6.2$\pm$0.3&5.4$\pm$0.4\\
                &&nthcomp&      $\Gamma$&2.01$\pm$0.01&1.95$\pm$0.01&2.03$\pm$0.01&1.92$\pm$0.01&1.93$\pm$0.01\\
                &&&kT$_{\rm e}$&>85&>40&>140&>60&>25\\
                &&&Norm ($\times$10$^{-4}$)&3.0$\pm$0.6&3.8$\pm$0.6&2.0$\pm$1.0&0.3$\pm$0.1&4.9$\pm$1.0\\
                &&relxillcp&r$_{\rm in}$ (r$_{\rm g}$)&2.7$\pm$0.6&>1.8&3.2$\pm$0.6&2.5$^{+0.5}_{-1.0}$&4.0$^{+3.1}_{-1.1}$\\
                &&&i$\dagger$ (deg)&69$\pm$1&&&&\\
                &&&$q\dagger$&>9.5&&&&\\
                &&&$\log \xi \dagger$&2.72$\pm$0.04&&&&\\
                &&&A$_{\rm Fe} \dagger$&1.55$\pm$0.25&&&&\\
                &&&Norm ($\times$10$^{-6}$)&6.2$\pm$0.5&6.3$\pm$0.9&7.3$\pm$0.7&4.0$\pm$0.2&3.3$\pm$0.4\\
                &&xillvercp&$\log \xi \dag$&0.15$\pm$0.12&&&&\\ 
                &&&Norm($\times$10$^{-4}$)&2.4$\pm$0.2&1.8$\pm$0.1&2.3$\pm$0.2&1.4$\pm$0.3&1.5$\pm$0.2\\
                \hline
                lamp post&1915/1567&diskbb&     T$_{\rm bb}$ ($\times10^{-3}$ eV)&3.5$\pm$0.5&&&& \\
                &&&Norm ($\times10^{9})$ &5.6$\pm$0.4&6.1$\pm$0.4&6.4$\pm$0.5&5.8$\pm$0.4&5.1$\pm$0.4\\
                &&nthcomp       &$\Gamma$&2.04$\pm$0.01&1.99$\pm$0.01&2.06$\pm$0.01&1.95$\pm$0.02&1.96$\pm$0.01\\
                &&&kT$_{\rm e}$&>30&>25&>120&>43&>17\\
                &&&Norm ($\times$10$^{-3}$)&1.47$\pm$0.06&1.43$\pm$0.07&1.50$\pm$0.08&0.58$\pm$0.06&1.02$\pm$0.06\\
                &&relxilllpcp&r$_{\rm in}$ (r$_{\rm g}$)&>2.5&>3&2.7$^{+0.8}_{-0.6}$&>2&3$^{+4.9}_{-1.3}$\\
                &&&i$\dagger$ (deg)&84$\pm$1&&&&\\
                &&&$h\dagger$&6.3$\pm$0.5&&&&\\
                &&&$\log \xi \dagger$&2.88$\pm$0.04&&&&\\
                &&&A$_{\rm Fe} \dagger$&0.55$\pm$0.05&&&&\\
                &&&Norm ($\times$10$^{-5}$)&5.0$\pm$1.6&3.8$\pm$0.6&5.8$\pm$1.0&2.7$\pm$0.5&2.8$\pm$1.2\\
                &&xillvercp&$\log \xi \dag$&0.03$\pm$0.01&&&&\\ 
                &&&Norm($\times$10$^{-4}$)&2.3$\pm$0.1&1.9$\pm$0.1&2.3$\pm$0.1&1.5$\pm$0.1&1.5$\pm$0.1\\
                \hline
                
                high density&1787/1567&diskbb&  T$_{\rm bb}$ ($\times10^{-3}$ eV)&3.8$\pm$0.5&&&& \\
                &&&Norm ($\times10^{9})$ &4.0$\pm$0.4&4.4$\pm$0.3&4.5$\pm$0.3&4.3$\pm$0.3&3.7$\pm$0.4\\
                &&nthcomp       &$\Gamma$&2.03$\pm$0.01&1.98$\pm$0.01&2.05$\pm$0.01&1.93$\pm$0.02&1.94$\pm$0.01\\
                &&&kT$_{\rm e}$&>85&>35&>105&>40&>22\\
                &&&Norm ($\times$10$^{-4}$)&1.85$\pm$0.05&1.78$\pm$0.03&1.94$\pm$0.03&0.82$\pm$0.06&1.22$\pm$0.04\\
                
                &&relxillD&r$_{\rm in}$ (r$_{\rm g}$)&>2&>1.7&>1.7&>1.6&>2.2\\
                &&&$q$&2.15$\pm$0.10&&&&\\
                &&&i$\dagger$ (deg)&84$\pm$1&&&&\\
                &&&$\log\rho\dagger$&18.05$\pm$0.03&&&&\\
                &&&$\log \xi \dagger$&1.3$\pm$0.1&&&&\\
                &&&A$_{\rm Fe} \dagger$&0.70$\pm$0.08&&&&\\
                &&&Norm ($\times$10$^{-6}$)&1.63$\pm$0.2&1.41$\pm$0.2&1.67$\pm$0.2&1.1$\pm$0.2&0.9$\pm$0.2\\
                &&xillvercp&$\log \xi \dag$&0.07$\pm$0.01&&&&\\ 
                &&&Norm($\times$10$^{-4}$)&1.6$\pm$0.1&1.3$\pm$0.2&1.6$\pm$0.1&0.98$\pm$0.12&1.2$\pm$0.1\\
                
        \end{tabular}
\end{table*}

Different \textit{Relxill} versions have been published \citep[e.g.][]{Daus16} and each of them accounts for a specific physical and geometrical corona-disc scenario. Therefore, we further tested the relativistic reflection scenario, replacing \textit{relxillcp} with \textit{relxilllpcp,} which accounts for a lamp-post corona-SMBH geometry. In this model, it is possible to fit the height ($h$, in units of gravitational radii r$_{\rm g}$) of the hot corona above the accretion disc, which is split into multiple zones, with each of them seeing a different incident spectrum due to relativistic energy shifts of the nuclear continuum. The fitting procedure is similar to the one described for the case of \textit{relxillcp,} and we calculated the coronal height above the accretion disc by tying its value among the pointings. The fit returns $h$=6.3$\pm$0.5 r$_{\rm g,}$ but the corresponding statistical quality is worse with respect to the previous test as we find $\chi^2$/d.o.f.=1915/1567. The primary continuum is characterised by changes in the source spectral shape, and the obtained values are slightly steeper with respect to those previously computed with \textit{relxillcp}. A sub-solar value of A$_{\rm Fe}$=0.55$\pm$0.05 is obtained. The ionisation parameter for the relativistic reflection component is $\log(\xi/{\rm erg~s^{-1}~cm })$=2.88$\pm$0.04, while for \textit{xillvercp} we get $\log(\xi/{\rm erg~s^{-1}~cm }$)=0.03$\pm$0.01~. The current fit returns a well-constrained disc inclination, though its value is extreme at $i$=84$\pm$1 deg. As a last attempt, we left the corona height free to vary between the
observations. However, no statistically significant variability of $h$ is
found.\\
\indent Finally, we tested the high-density version of \textit{relxill,}  which is \textit{relxillD} \citep[][high density model in Table~\ref{relativistic}]{Daus16}. Such a model allows for the density of the accretion disc ($\log\rho/$cm$^{-3}$) to vary in the 15-19 range, where the standard version of the reflection code assumes this quantity to be $\log\rho/$cm$^{-3}$=15. However, in its current version, \textit{relxillD} does not allow for a variable high -energy cut off that is assumed to be frozen at 300 keV. Once \textit{relxillcp} had been replaced with its high-density counterpart, we fitted $\rho,$ tying its value between the different observations, while the other parameters were treated as in case of \textit{relxillcp}. The fit to the data returns a disc-density value of $\log$($\rho$/cm$^{-3}$)=18.05$\pm$0.03, the other parameters being marginally affected by this new model and consistent within the errors with what was previously obtained, with the exception of the ionisation parameter for the relativistic component found to be $\log(\xi/{\rm\rm erg s^{-1}~cm})$=1.3$\pm$0.1: much smaller than in previous fits. This model is better on statistical grounds ($\chi^2$=1789 for 1567 d.o.f.) than the previously tested reflection models. Also, for this modelling the iron abundance is sub-solar (A$_{\rm Fe}$=0.70$\pm$0.08). In this case, the corona emissivity is not compatible with what was previously estimated by \textit{relxillcp} since the fit gives $q$=2.15$\pm$0.10 for this parameters, while an extreme disc inclination of $i$=83$\pm$1 deg is found.\\
\indent The best fit parameters for all the models tested in this section are reported in Table \ref{relativistic}. 
In accordance with our tests, \textit{relxillD} provides the most satisfactory fit among the relativistic reflection models, though the emerging scenario is characterised by a large disc inclination.

\section{Summary and conclusions}
In this section, we first summarise the main results of the present analysis, and then propose a possible scenario.
\subsection{Results}
We presented the UV-to-X-ray spectral characterisation of Mrk 359, the first discovered NLSy1. Data belong to a monitoring campaign of 5$\times$ 50 ks \textit{XMM-Newton-NuSTAR} simultaneous exposures taken every $\sim$2 days, see Table 1.
In the following, we highlight the main results of this work:
\begin{itemize}

\item The Seyfert galaxy Mrk 359 varied on different timescales and spectral bands as shown in Fig.~\ref{lc}. In fact, amplitude changes up to a factor of $\lesssim$3 occurred in the \textit{XMM-Newton} 0.3-2 and 2-10 keV bands, as well as in the harder \textit{NuSTAR} bandpass. Interestingly, the \textit{XMM-Newton} light curves computed for the first three observations are a factor of $\sim$3 higher with respect to the fourth and fifth pointings, and a similar behaviour seems to characterise the \textit{OM} UVM2 and UVW2 filters (see Fig.\ref{om}). 
The suggestive correlation between UV photons and the soft X-ray band is shown in Fig~\ref{Xuv}, and this is in agreement with what is expected from the two-corona model. Hardness ratios between the 0.3-2 and 2-10 keV bands are weakly variable within the same pointing, though spectral changes are observed between the different exposures. The light curves shown in Fig.~\ref{lc} are consistent with the 'softer when brighter' behaviour \citep[e.g.][]{Bian09,Sobolewska09,Serafinelli17}, meaning that harder spectral states correspond to lower fluxes.
\item The F$_{\rm var}$ spectra are in agreement with two variability states Mrk 359 underwent during the monitoring; observations 1, 2, and 3 are indeed characterised by smaller amplitude variability with respect to observations 4 and 5. Though limited in terms of S/N, the $F_{\rm var}$ spectra may suggest that about the same amount of variability occurs over the different energy intervals. The only exception may be observations 4 and 5, for which the soft X-ray band seems to have larger F$_{\rm var}$ values.
\item We extracted the \textit{EPIC-pn} light curves in different energy intervals (0.5-2, 2.5-5 and 8-10 keV) to investigate whether these light curves vary in concert. Such bands, in fact, are likely dominated by three different spectral components: the soft-excess, the primary continuum, and the reflected emission. Based on the Pearson cross-correlation analysis, we find that the strong correlation P$_{\rm cc}$=0.91, P(r<0)<10$^{-22}$ holds between the 0.5-2 and 2.5-5 light curves, while weaker correlation (also in term of statistics) characterises the light curves in the soft X-ray band and the 8-10 keV one (P$_{\rm cc}$=0.42, P(r<0)<10$^{-6}$). Such a result is not consistent with the scenario in which the soft excess is due to relativistic lines blurring, since the reflection component would be present in both energy bands and a correlation would be expected.
\item Using the 2-10 keV band, we quantified the amount of variability in Mrk 359 computing the normalised excess variance (see Sect. 3 for details) that we also used it to estimate Mrk 359 SMBH mass. Following the prescriptions by \cite{Ponti12}, we found M$_{\rm BH}$=(3.6$\pm$1.4)$\times$10$^6$ M$_{\odot}$.
\item The 3-79 keV spectra of Mrk 359 are well described by a power-law-like emission with a $\Gamma_{\rm hc}$ varying between 1.75 and 1.88 and a high-energy cut off E$_{\rm c}$>220 keV, see Table 3. This primary continuum emission is accompanied by a neutral Fe K$\alpha$ emission line (EW$\sim$90 eV). Such a feature has a width consistent with zero throughout the campaign (see Fig.~\ref{width}), with no variability observed. \textit{NuSTAR} data show a more constant behaviour during the campaign and are characterised by the Compton hump shown in Fig.~\ref{soft_excess}. The flux of the reprocessed emission is constant during the monitoring, and the origin of this component is consistent with Compton-thick and cold matter \citep[as commonly found in other Seyfert galaxies, see e.g.][]{Bian09,Cappi2016}.\\
\item We fit the broadband UV-to-X-ray emission spectrum of Mrk 359 with two different models: a two-corona model and relativistic reflection. The first model is favoured on statistical grounds and is characterised by a hot component (kT$_{\rm hc}$>55 keV, $\tau_{\rm tc }$<4.3), which accounts for the primary continuum, and a warmer one (kT$_{\rm wc}\sim0.3$ keV, $20\leq\tau_{\rm WC }\leq29$) responsible for the UV and soft X-ray emission. The soft-excess can be ascribed to this latter component. Weak/strong spectral variability are observed on the hot/warm phase, as shown in Fig.~\ref{cont}.\\
\indent         The relativistic reflection code accounting for a variable density of the accretion disc, \textit{relxillD}, provides the better fit among the various \textit{relxill} versions (see also Fig.~\ref{refl}),  though its $\chi^2$/d.o.f. is significantly larger than that of the two-corona model. It is also worth mentioning that the disc inclination angles obtained with all the relativistic reflection models are fairly extreme, and no evidence of such a large inclination exists for Mrk 359. This also weakens a possible relativistic origin of the soft excess in Mrk 359. \\

\end{itemize}
\subsection{Two coronae in Mrk 359 and a possible physical scenario}
\begin{figure}
        \centering      
        \includegraphics[width=0.48\textwidth]{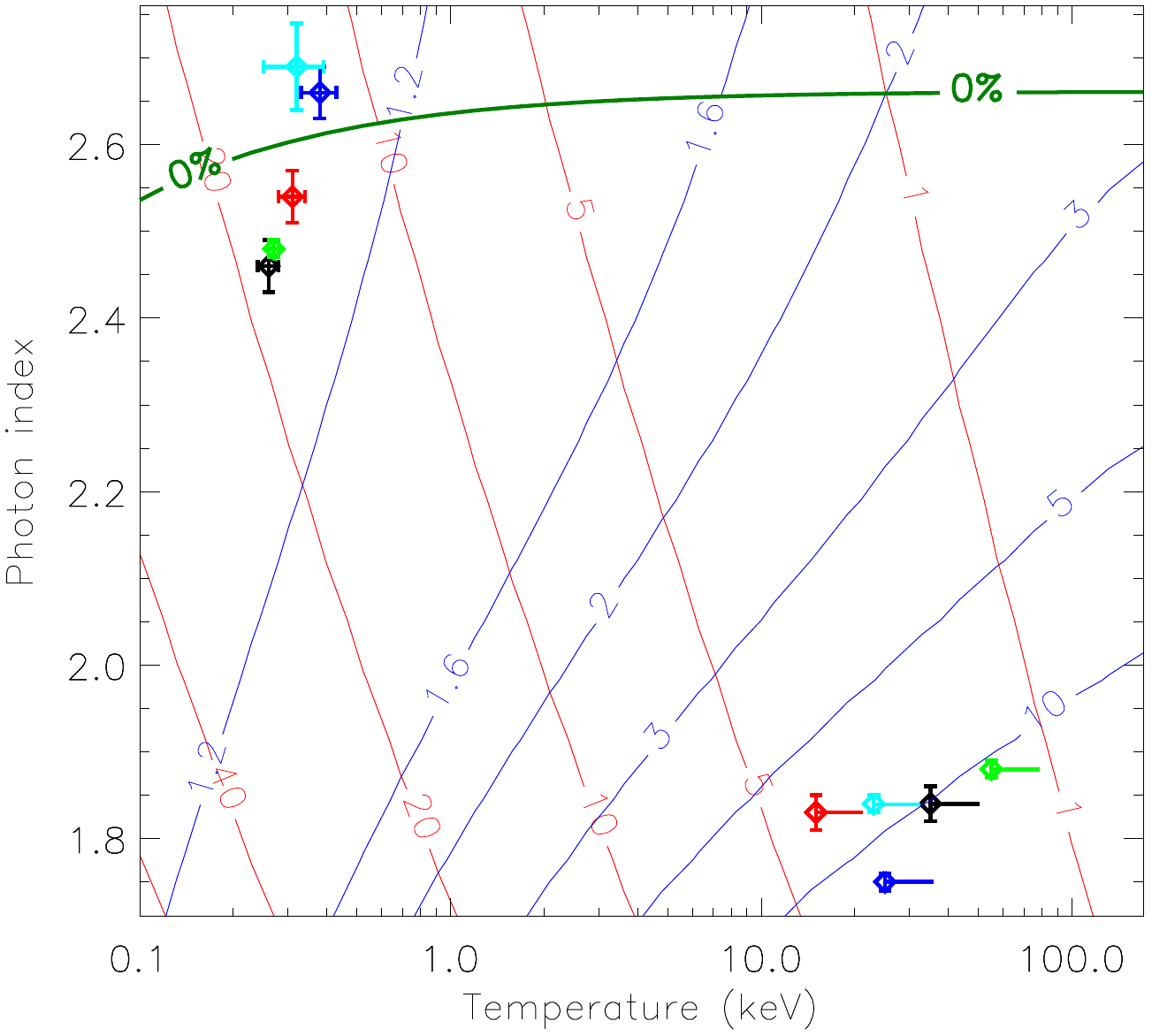}      
        \caption{\small{For each observation, $\Gamma$ and $kT$ of the two coronae are shown.  The solid lines refer to the corona optical depth $\tau$ (red), the amplification factor A (in blue) and the minimal fraction (in \%) of disc intrinsic emission (dark green). Warm corona points lie both below (observations 1, 2, and 3) and above (observations 4 and 5) the zero intrinsic-disc emission line, and this agrees with a change of radiative equilibrium for this component \citep[see][for a detailed discussion]{Petrucci2020}. \label{popcont}}}
\end{figure}

\begin{figure}
        \centering
        \includegraphics[width=0.48\textwidth]{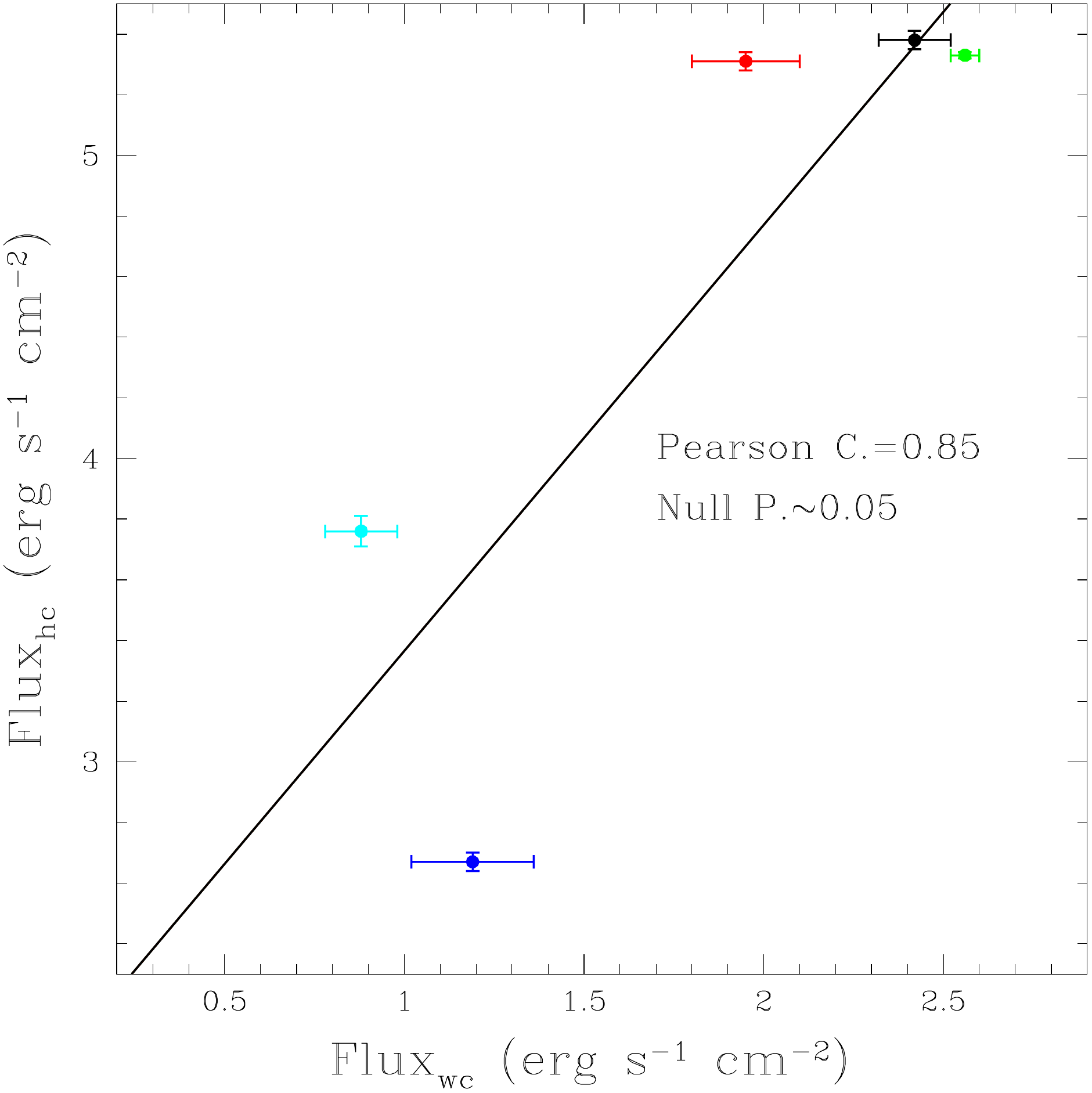}      
        \caption{\small{Correlation between flux of the hot Comptonisation component in the 2-10 keV band and the soft one estimated in the 0.3-2 keV band. Fluxes are in units of 10$^{-12}$ erg s$^{-1}$ cm$^{-2.}$ \label{plot}}}
\end{figure}

Our analysis shows that the two-corona model is the one that best fits the data, so we discuss it in some detail. The evolution of the physical parameters (optical depth, temperature, photon index) of the two coronae can be used to derive a tentative scenario of the spectral behaviour of the source observed during the campaign.\\
\indent Following \citet{Petrucci2020}, we can estimate the amplification factors for the two coronae. We find A$_{\rm wc}$<1.1 and A$_{\rm hc}\sim$7 for the warm and hot coronas, respectively. The large amplification factor of the hot corona is commonly found in AGNs and is consistent with a scenario in which the seed photons that are Comptonised represent only a small fraction of the photons produced by the accretion disc. This supports geometries like a patchy corona, or a corona inside a truncated disc, or a corona on the SMBH axis (i.e. the so-called lamp-post model). On the other hand, in the case of the warm corona, the value of the temperature and spectral index agree with an amplification factor of $\sim$1.1, which is consistent with the corona covering a quasi-passive accretion disc. This is shown in Fig.~\ref{popcont}, where we report the hot and warm corona's properties in the spectral photon index versus the temperature plane. This figure is similar to the one discussed in \citet[][Fig. 1]{Petrucci18}, but it was corrected in \citet[][Fig. 8]{Petrucci2020}. In fact, concerning the warm corona, the five observations are located in two slightly different regions. Observations 1, 2, and 3 have harder warm-corona spectra and are located below the passive disc line, while observations 4 and 5 are above this line (due to their softer warm-corona spectra). This clearly suggests a change of the warm-corona radiative equilibrium during the beginning and the end of the monitoring. Interestingly, Observations 1, 2, and 3 have a higher flux than observations 4 and 5, indicating that this change of the warm-corona radiative equilibrium could go with a change of the source flux state.\\
\indent A tentative scenario to explain the observed spectral variability of Mrk 359 could be the following: assuming a hot corona/truncated-disc geometry where the outer disc is covered by a warm corona, the warm corona's photons may then enter and cool the hot corona through Comptonisation. This is indeed supported by the suggestive correlation between the hard and soft X-ray flux reported in Fig.~\ref{plot}. We may also assume that the two flux regimes observed during the campaign are the result of a change in the accretion rate $\dot{m}$ (i.e. the accretion rate slightly decreases between the beginning and the end of the monitoring), and that this change of $\dot{m}$ goes along with a change of the truncation radius R$_{\rm tr}$. In fact, such a relation between $\dot{m}$ and R$_{\rm tr}$ is expected in the disc evaporation model developed by, for example, \citet[][]{Meyer2000} and \citet{Ronza2000} \citep[see also][for a review]{Czerny2018}. In this approach, this inner hot corona is fed by matter from the outer thin disc, which evaporates from the cool layers underneath, the evaporation process resulting from thermal instabilities led by the sharp temperature gradient between the hot corona and the outer disc. The warm corona, at the surface of the accretion disc, could be a natural product of this evaporation process. Now the strong dependence of the cooling efficiency on density leads to an anti-correlation between the accretion rate and the truncated radius.
So, in this framework, the apparent decrease of the accretion rate during the campaign of Mrk 359 would go along with an increase of the truncation radius. This would mean that fewer photons coming from the outer accretion flow will enter and cool the hot corona, decreasing the cooling process, and resulting in a hardening of its spectrum as observed (see Fig.~\ref{gamsh}, bottom). On the other hand, the decrease in the accretion rate and increase in the truncation radius will naturally decrease the intrinsic warm corona heating power, as well as the illumination from the hot corona. As a consequence, the warm-corona spectrum would soften in agreement with the observations (see Fig.~\ref{gamsh}, top).
We note that this approach supposes small and local variation timescales of the accretion rate and the truncation radius. This is similar to the case of HE 1143-1810  \citep{Ursini2019}, an AGN that behaves similarly to Mrk 359 in X-rays.  These timescales are much shorter (< day) than the viscous timescale needed to reach the ISCO (innermost stable circular orbit) from the truncation radius, and this agrees quite well with the variability timescales observed in our monitoring.\\
\indent While this scenario appears to be in qualitative agreement with the observed spectral behaviour of Mrk 359, it is not entirely satisfactory. For example, it does not explain the apparent increase in the warm-corona temperature during the campaign (see Fig.\ref{cont}), suggesting that other parameters are also playing a role. In any case, better constraints on the truncation radius are needed here to confirm or rule out the present interpretation, and high-resolution instruments that allow highly statistical analyses like those of the Athena mission will be crucial in this respect.


\begin{acknowledgements}
        
RM warmly thanks Eleonora Bianchi for being helpful during his time in Grenoble and Fausto Vagnetti, Barbara De Marco \& Gabriele Matzeu for useful comments. RM also acknowledges Fondazione Angelo Della Riccia for financial support and Universit\'e Grenoble Alpes and the high energy SHERPAS group for welcoming him at IPAG. SB acknowledges financial support from ASI under grants ASI-INAF I/037/12/0 and n. 2017-14-H.O. POP and MCl acknowledges financial support from the CNRS PNHE and from the CNES french agency. This  work  is  based  on  observations obtained with: the NuSTAR mission,  a  project  led  by  the  California  Institute  of  Technology,  managed  by  the  Jet  Propulsion  Laboratory  and  funded  by  NASA; XMM-Newton,  an  ESA  science  mission  with  instruments  and  contributions  directly funded  by  ESA  Member  States  and  the  USA  (NASA). ADR \& MC acknowledge financial contribution from the agreement ASI-INAF n.2017-14-H.O. Part of this work is based on archival data, software or online services provided by the Space Science Data Center - ASI.
\end{acknowledgements}

\thispagestyle{empty}
\bibliographystyle{aa}
\bibliography{mrk359_accepted.bib}

\end{document}